\documentstyle[12pt]{article}
\catcode`\@=11
\@addtoreset{equation}{section}

\def\theequation{\thesection.\arabic{equation}}

\newtoks\@stequation

\def\subequations{\refstepcounter{equation}%
  \edef\@savedequation{\the\c@equation}%
  \@stequation=\expandafter{\theequation}
  \edef\@savedtheequation{\the\@stequation}
  \edef\oldtheequation{\theequation}%
  \setcounter{equation}{0}%
  \def\theequation{\oldtheequation\alph{equation}}}

\def\endsubequations{\setcounter{equation}{\@savedequation}%
  \@stequation=\expandafter{\@savedtheequation}%
  \edef\theequation{\the\@stequation}\global\@ignoretrue
  \vspace*{-12pt} \\}

\def\hybrid{\topmargin -20pt	\oddsidemargin 0pt
	\headheight 0pt	\headsep 0pt
	\textwidth 6.25in	
	\textheight 9.5in	
	\marginparwidth .875in
	\parskip 5pt plus 1pt	\jot = 1.5ex}

\def\baselinestretch{1.2}

\catcode`\@=11

\def\marginnote#1{}
\newcount\hour
\newcount\minute
\newtoks\amorpm
\hour=\time\divide\hour by60
\minute=\time{\multiply\hour by60 \global\advance\minute by-\hour}
\edef\standardtime{{\ifnum\hour<12 \global\amorpm={am}%
	\else\global\amorpm={pm}\advance\hour by-12 \fi
	\ifnum\hour=0 \hour=12 \fi
	\number\hour:\ifnum\minute<10 0\fi\number\minute\the\amorpm}}
\edef\militarytime{\number\hour:\ifnum\minute<10 0\fi\number\minute}
\def\draftlabel#1{{\@bsphack\if@filesw {\let\thepage\relax
   \xdef\@gtempa{\write\@auxout{\string
      \newlabel{#1}{{\@currentlabel}{\thepage}}}}}\@gtempa
   \if@nobreak \ifvmode\nobreak\fi\fi\fi\@esphack}
	\gdef\@eqnlabel{#1}}
\def\@eqnlabel{}
\def\@vacuum{}
\def\draftmarginnote#1{\marginpar{\raggedright\scriptsize\tt#1}}

\def\draft{\oddsidemargin -.2truein
	\def\@oddfoot{\sl preliminary draft \hfil
	\rm\thepage\hfil\sl\today\quad\militarytime}
	\let\@evenfoot\@oddfoot	\overfullrule 3pt
	\let\label=\draftlabel
	\let\marginnote=\draftmarginnote
   \def\@eqnnum{(\theequation)\rlap{\kern\marginparsep\tt\@eqnlabel}%
\global\let\@eqnlabel\@vacuum}  }

\def\preprint{\twocolumn\sloppy\flushbottom\parindent 2em
	\leftmargini 2em\leftmarginv .5em\leftmarginvi .5em
	\oddsidemargin -.5in	\evensidemargin -.5in
	\columnsep .4in	\footheight 0pt
	\textwidth 10.in	\topmargin  -.4in
	\headheight 12pt \topskip .4in
	\textheight 6.9in \footskip 0pt
	\def\@oddhead{\thepage\hfil\addtocounter{page}{1}\thepage}
	\let\@evenhead\@oddhead	\def\@oddfoot{}	\def\@evenfoot{} }

\def\titlepage{\@restonecolfalse\if@twocolumn\@restonecoltrue\onecolumn
     \else \newpage \fi \thispagestyle{empty}\c@page\z@
	\def\thefootnote{\fnsymbol{footnote}} }

\def\endtitlepage{\if@restonecol\twocolumn \else \newpage \fi
	\def\thefootnote{\arabic{footnote}}
	\setcounter{footnote}{0}}  

\catcode`@=12
\relax

\def\figcap{\section*{Figure Captions\markboth
	{FIGURECAPTIONS}{FIGURECAPTIONS}}\list
	{Figure \arabic{enumi}:\hfill}{\settowidth\labelwidth{Figure
999:}
	\leftmargin\labelwidth
	\advance\leftmargin\labelsep\usecounter{enumi}}}
 \relax
\def\tablecap{\section*{Table Captions\markboth
	{TABLECAPTIONS}{TABLECAPTIONS}}\list
	{Table \arabic{enumi}:\hfill}{\settowidth\labelwidth{Table
999:}
	\leftmargin\labelwidth
	\advance\leftmargin\labelsep\usecounter{enumi}}}
 \relax
\def\reflist{\section*{References\markboth
	{REFLIST}{REFLIST}}\list
	{[\arabic{enumi}]\hfill}{\settowidth\labelwidth{[999]}
	\leftmargin\labelwidth
	\advance\leftmargin\labelsep\usecounter{enumi}}}
 \relax
\def\R{{\rm I\!R}}

\def\one{{\mathchoice {\rm 1\mskip-4mu l} {\rm 1\mskip-4mu}
{\rm 1\mskip-4.5mu l} {\rm 1\mskip-5mu l}}}
\def\Q{{\mathchoice
{\setbox0=\hbox{$\displaystyle\rm Q$}\hbox{\raise 0.15\ht0\hbox to0pt
{\kern0.4\wd0\vrule height0.8\ht0\hss}\box0}}
{\setbox0=\hbox{$\textstyle\rm Q$}\hbox{\raise 0.15\ht0\hbox to0pt
{\kern0.4\wd0\vrule height0.8\ht0\hss}\box0}}
{\setbox0=\hbox{$\scriptstyle\rm Q$}\hbox{\raise 0.15\ht0\hbox to0pt
{\kern0.4\wd0\vrule height0.7\ht0\hss}\box0}}
{\setbox0=\hbox{$\scriptscriptstyle\rm Q$}\hbox{\raise 0.15\ht0\hbox to0pt
{\kern0.4\wd0\vrule height0.7\ht0\hss}\box0}}}}
\def\C{{\mathchoice
{\setbox0=\hbox{$\displaystyle\rm C$}\hbox{\hbox to0pt
{\kern0.4\wd0\vrule height0.9\ht0\hss}\box0}}
{\setbox0=\hbox{$\textstyle\rm C$}\hbox{\hbox to0pt
{\kern0.4\wd0\vrule height0.9\ht0\hss}\box0}}
{\setbox0=\hbox{$\scriptstyle\rm C$}\hbox{\hbox to0pt
{\kern0.4\wd0\vrule height0.9\ht0\hss}\box0}}
{\setbox0=\hbox{$\scriptscriptstyle\rm C$}\hbox{\hbox to0pt
{\kern0.4\wd0\vrule height0.9\ht0\hss}\box0}}}}

\font\fivesans=cmss10 at 4.61pt
\font\sevensans=cmss10 at 6.81pt
\font\tensans=cmss10
\newfam\sansfam
\textfont\sansfam=\tensans\scriptfont\sansfam=\sevensans\scriptscriptfont
\sansfam=\fivesans

\def\Z{Z\!\!\! Z}

\mathchardef\endbar="375

\makeatletter
\newcounter{pubctr}
\def\publist{\@ifnextchar[{\@publist}{\@@publist}}
\def\@publist[#1]{\list
	{[\arabic{pubctr}]\hfill}{\settowidth\labelwidth{[999]}
	\leftmargin\labelwidth
	\advance\leftmargin\labelsep
	\@nmbrlisttrue\def\@listctr{pubctr}
	\setcounter{pubctr}{#1}\addtocounter{pubctr}{-1}}}
\def\@@publist{\list
	{[\arabic{pubctr}]\hfill}{\settowidth\labelwidth{[999]}
	\leftmargin\labelwidth
	\advance\leftmargin\labelsep
	\@nmbrlisttrue\def\@listctr{pubctr}}}
 \relax
\makeatother
\newskip\humongous \humongous=0pt plus 1000pt minus 1000pt
\def\caja{\mathsurround=0pt}
\def\eqalign#1{\,\vcenter{\openup1\jot \caja
	\ialign{\strut \hfil$\displaystyle{##}$&$
	\displaystyle{{}##}$\hfil\crcr#1\crcr}}\,}
\newif\ifdtup

\relax
\hybrid
\def\a{\alpha}
\def\b{\beta}
\def\g{\gamma}  
\def\d{\delta}

\def\n{\nu}

\def\ps{\psi}
\def\r{\rho}

\def\s{\sigma}

\def\et{\eta}
\def\L{\Lambda}

\def\D{\Delta}
\def\bz{\bar{z}}
\def\bd{\bar{d}}

\def\bw{\bar{w}}
\def\bv{\bar{v}}
\def\bq{\bar{q}}

\def\by{\bar{y}}
\def\bs{\bar{s}}

\def\bX{\bar{X}}
\def\bN{\bar{N}}

\def\bA{\bar{A}}
\def\bL{\bar{L}}
\def\bJ{\bar{J}}

\def\bQ{\bar{Q}}
\def\pG{G_+}
\def\Gp{G_+}
\def\Gm{G_-}
\def\bG{G_-}
\def\bF{\bar{F}}
\def\bC{\bar{C}}

\def\J{J}
\def\F{{\cal F}}

\def\G{{\cal G}}

\def\T{{\cal T}}
\def\bT{\bar{T}}
\def\C{{\cal C}}
\def\cO{{\cal O}}

\def\B{{\cal B}}

\def\pa{\partial}
\def\bpa{\bar{\partial}}
\def\ve{\vert}

\def\ra{\rightarrow}
\def\lra{\leftrightarrow}
\def\ti{\times}
\def\oti{\otimes}

\def\xx{\hbox{ }^*_*}
\def\gp{g_+}
\def\gm{g_-}

\def\ok{\cO (k^{-2} )}
\def\oko{\cO (k^{-1} )}
\def\okt{\cO (k^{-3/2} )}
\def\okh{\cO (k^{1/2} )}

\def\un{\underline}
\def\ni{\noindent}

\def\thefootnote{\fnsymbol{footnote}}
\def\be{\begin{equation}}
\def\ee{\end{equation}}
\def\bs{\begin{subequations}}
\def\es{\end{subequations}}
\def\ben{\begin{enumerate}}
\def\een{\end{enumerate}}

\def\G{{\cal G}}

\def\vs{\vskip}

\def\rd{{\rm d}}

\def\ni{\noindent}

\def\ed{\end{document}}

\def\bibtem#1{\bibitem{#1} }
\def\cit#1{\cite{#1}}

\def\sp{\quad, \quad}
\def\pe{\quad . }

\begin{document}
\begin{titlepage}
\begin{center}

\hfill UCB-PTH-96/41 \\ 
\hfill LBNL-39438 \\  
\hfill CPTH-S467.1096 \\
\hfill hep-th/9610081  \\
\vskip .2in

{\large \bf  
New Semiclassical Nonabelian Vertex Operators for  \\  
Chiral and Nonchiral WZW Theory}
\vskip .4in
{\bf M.B. Halpern}
\footnote{e-mail address: halpern@theor3.lbl.gov} 
\\
\vskip .15in
{\em 
Department of Physics,  
 University of California \\
and \\   
Theoretical Physics Group, Lawrence Berkeley National Laboratory\\
 Berkeley, CA 94720, USA}  
\vskip .2in
and
\vskip .15in
{\bf N.A. Obers}
\footnote{Address after Oct. 1, 1996: Theory Division CERN, CH-1211 Geneva 23,
Switzerland; \newline \indent $\;\;$e-mail address: obers@mail.cern.ch}
\vskip .1in
{\em Centre de Physique 
Th\'eorique\footnote{Laboratoire Propre du CNRS UPR A.0014}, 
Ecole Polytechnique \\
F-91128 Palaiseau,
FRANCE } 
\end{center}

\vskip .3in

\begin{center} {\bf ABSTRACT } \end{center}
\begin{quotation}\noindent
We supplement the discussion of Moore and Reshetikhin and others
by finding new   
semiclassical nonabelian vertex operators for the chiral, antichiral and 
nonchiral primary fields of WZW theory. These new nonabelian vertex operators
are the natural generalization of the familiar abelian vertex 
operators: They
involve only the representation matrices of Lie $g$, the currents of affine
$(g \ti g)$ and certain chiral and antichiral zero modes, and they reduce to
the abelian vertex operators in the limit of abelian algebras.
Using the new constructions, we also discuss semiclassical operator product 
expansions, braid relations and relations to the known form of the
semiclassical affine-Sugawara conformal blocks.
\end{quotation}
 \vskip 0.7cm
 \end{titlepage}

\vfill
\eject
\def\baselinestretch{1.2}
\baselineskip 16 pt

\setcounter{equation}{0}

\section{Introduction} 

Affine Lie algebra [1,2] is the basis of a very large set of conformal
field theories
called the affine-Virasoro constructions [3,4] which include the
affine-Sugawara
constructions [2,5,6,7], the coset constructions [2,5,8] 
 and the irrational conformal field theories [3,9,10].  
Among these, the simplest theories are the affine-Sugawara
constructions and their corresponding WZW actions [11,12], 
which have often served as a testing
ground for new ideas in conformal field theory. 
See Ref.\cit{10} for a more detailed history of
affine Lie algebra and the affine-Virasoro constructions.

Vertex operator constructions (see for example 
[13-22])  
are explicit realizations (using the familiar abelian vertex operators 
\cit{23}) of the fermions, currents and primary fields of affine
Lie algebras and conformal field theories. The first vertex operator 
constructions [13-15] were the vertex operator constructions of world-sheet
fermions and  level one of untwisted  $SU(n)$, which was also the 
first construction of current-algebraic internal
 symmetry from compactified dimensions on the string. The
generalization [17,18] 
of this construction to level one of simply laced $g$  
plays a central role in the formulation of the heterotic string \cit{24}. 
More generally, the vertex operator constructions may be divided into the
explicitly unitary constructions [13-18] and the
constructions of (bosonized) Wakimoto type [19-22], which must 
be projected onto unitary subspaces.

In this paper, we will supplement the discussion of Moore and Reshetikin 
\cit{25} and others [26-41]
by finding a new explicit semiclassical (high-level) realization of the
chiral, antichiral and nonchiral primary fields of WZW theory. 
The realization is  obtained by semiclassical solution of known [26,25] 
operator differential equations for the chiral primary fields
(so that no unitary projection is needed), and the results are
recognized as the semiclassical form of new nonabelian vertex operators which
are the natural generalization of the familiar abelian vertex operator: 
In particular, the new nonabelian vertex operators
 involve only the
representation matrices of Lie $g$, the currents of affine $(g \ti g)$ and
certain chiral and antichiral zero
modes, and they reduce to the abelian vertex operators in the limit of abelian 
algebras.

 A central feature of the construction is the identification of the chiral
and antichiral zero modes which, thru the semiclassical order we have
studied, are seen
to carry the full action of the quantum group. We are also able to identify
the classical
limit of the nonchiral product of the zero modes as the classical group element.

As applications of our construction, we compute the
semiclassical OPE's of
all the primary fields and compare the averages  of the primary fields to
the known \cit{42} forms
of the semiclassical
affine-Sugawara conformal blocks and WZW correlators. The relation of the
construction
to semiclassical crossing and braiding is also discussed.

\section{Affine Lie Algebra and WZW Theory} 

In this section we review some basic facts about affine Lie algebra 
[1,2] and 
the affine-Sugawara constructions [2,5,6,7] which provide the 
algebraic description of WZW theory \cit{12}.

We begin with the algebra of affine $(g \ti g)$, which consists of 
two commuting copies of affine $g$,
\bs
\be
[\J_a(m),\J_b(n) ] = i f_{ab}{}^c\J_c(m+n) + k m\et_{ab} \d_{m+n,0} 
\ee
\be
[ \bJ_a(m), \bJ_b(n) ] = i f_{ab}{}^c \bJ_c(m+n) + k m \et_{ab} \d_{m+n,0} 
\ee
\be
[\J_a(m), \bJ_b(n) ] = 0 \sp m,n \in \Z  
\sp a,b,c = 1 \ldots {\rm dim}\,g  
\ee
\es 
where $f_{ab}{}^c$ and $\et_{ab}$ are the structure constants and  
Killing metric of $g$ and $k$ is the level of the affine algebra. For 
simplicity we generally assume here that $g$ is compact, though most of the
statements below apply as well to the noncompact extensions of $g$. 
The affine vacuum state
$| 0 \rangle$ satisfies 
\be
\J_a (m \geq 0) |0 \rangle = 
\bJ_a (m \geq 0) |0 \rangle = 0 \pe  
\label{ava} \ee
In terms of these current modes, the local chiral and antichiral currents
are defined as
\be
\J_a (z) \equiv  \sum_{m \in \Z}\J_a (m) z^{-m-1} 
\sp
\bJ_a (\bz) \equiv \sum_{m \in \Z} \bJ_a (m) \bz^{-m-1} 
\ee
where $z$ is the complex Euclidian world-sheet coordinate and
$\bz$ is the complex conjugate of $z$.

We will also need the primary fields $g(\T,\bz,z)$ of affine $(g\ti g)$, which
transform under the current modes as
\bs
\be
[\J_a(m),  g(\T,\bz,z)_\a{}^\b ] = g(\T,\bz,z)_\a{}^\g z^m   (\T_a)_\g{}^\b  
\label{egc}
\ee
\be 
\; \;\; 
[\bJ_a(m),  g(\T,\bz,z)_\a{}^\b ] =-\bz^m  (\T_a)_\a{}^\g  g(\T,\bz,z)_\g{}^\b  
\ee
\be
[\T_a ,\T_b] = i f_{ab}{}^c \T_c 
\sp \a,\b = 1 \ldots {\rm dim}\,\T
\ee
\label{ega} \es 
where $\T$ is a matrix irrep of $g$. The primary fields $g(\T,\bz,z)$ and the
currents $\J,\bJ$ may be understood \cit{43} respectively as the (reduced) 
affine Lie group element and the (reduced) left- and right-invariant affine
Lie derivatives 
on the manifold of the affine Lie group. Acting on the affine vacuum,
the primary fields create the primary states $\ps (\T)$ of affine
$(g \ti g)$,
\bs
\be
\ps_\a{}^\b (\T) \equiv  g(\T,0,0)_\a{}^\b   |0 \rangle 
\ee
\be
\J_a (m \geq 0)\ps_\a{}^\b (\T) = 
\d_{m,0} \ps_\a{}^\g  (\T) (\T_a)_\g{}^\b  
\ee
\be
\;\;\; \bJ_a (m \geq 0)\ps_\a{}^\b (\T) = 
- \d_{m,0}(\T_a)_\a{}^\g   \ps_\g{}^\b  (\T) 
\ee
\label{aps} \es
which transform in irrep $\T \oti \bar{\T}$ of $(g\ti g)$. 
A coordinate-space representation of these states is given in Ref.\cit{43}.

The stress tensor of WZW theory is composed of the chiral and antichiral
affine-Sugawara constructions
\bs
\be
T_g(z) = L^{ab}_g \xx\J_a (z)\J_b (z) \xx 
= \sum_{m \in \Z} L_g(m) z^{-m-2} 
\ee
\be
\bar{T}_g(\bz) = L^{ab}_g \xx \bJ_a(\bz) \bJ_b (\bz) \xx 
= \sum_{m \in \Z} \bar{L}_g(m) \bz^{-m-2} 
\ee
\be
L^{ab}_g =  {\et^{ab}  \over 2 k + Q_g}    
\label{asl} \ee
\be
c_g = \bar{c}_g =   {2k \, {\rm dim}\,g \over 2 k + Q_g}    
\ee
\es
whose modes $L_g(m)$ and $\bL_g(m)$
satisfy two commuting Virasoro algebras with central charges $c_g$ and 
$\bar{c}_g$. Here, $Q_g$ is the quadratic Casimir of the adjoint and  
$L^{ab}_g$ is called the inverse inertia tensor of the
affine-Sugawara construction. The currents $\J,\bJ$ and the 
affine-primary fields
$g(\T,\bz,z)$ are also Virasoro primary fields under  $(T_g, \bar{T}_g)$ 
with conformal weights  $(1,0)$, $(0,1)$ and
$(\D^g(\T),\D^g(\T))$ respectively. The affine-Sugawara conformal 
weight $\D^g(\T)$ is given by  
\be
L^{ab}_g \T_a \T_b = \D^g(\T) \one \sp
\D^g (\T) = {Q(\T) \over 2k + Q_g} 
\label{acw} \ee 
with $Q(\T)$ the quadratic Casimir of $\T$.   
In what follows, the affine- and Virasoro-primary fields
$g(\T,\bz,z)$ are generally called the WZW primary fields.  

In the WZW action, the classical analogue of the WZW primary
field $g(\T,\bz,z)$ appears as the unitary Lie group element in irrep $\T$ of 
$g$.  We shall see below however that the WZW 
primary field $g(\T,\bz,z)$ is a unitary operator 
only in the extreme semiclassical limit, due to normal ordering in the quantum
theory.

\vs .4cm
\ni \un{Differential equations}
\vs .3cm

Because they are also primary under $(T_g,\bT_g)$, the WZW primary fields  
$g (\T,\bz,z)$ satisfy the operator relations 
\bs
\be
\pa  g(\T,\bz,z) =  [L_g(-1), g(\T,\bz,z)]
\sp  
\bpa  g(\T,\bz,z) =  [\bar{L}_g(-1), g(\T,\bz,z)]
\label{del} \ee
\be
L_g(-1) = 2 L^{ab}_g  \sum_{m \geq 0 }\J_a(-m-1)\J_b(m)
\sp 
\bar{L}_g(-1) = 2 L^{ab}_g  \sum_{m \geq 0 } \bJ_a(-m-1) \bJ_b(m)
\label{lex} \ee
\es
\vskip -.3cm \noindent  
and, using (\ref{lex}) in (\ref{del}), one finds the
partial differential equations (PDE's) for the WZW primary fields \cit{7}  
\bs
\be
\pa g(\T,\bz,z)_\a{}^\b  = 2 L^{ab}_g : g(\T,\bz,z)_\a{}^\g \J_a (z) : 
(\T_b)_\g{}^\b    
\label{cde} \ee
\be
\;\; \;\;\;\;\;  
\bpa g(\T,\bz,z)_\a{}^\b  = - 2 L^{ab}_g (\T_b)_\a{}^\g : g(\T,\bz,z)_\g{}^\b 
 \bJ_a (\bz) : \pe    
\label{ade} \ee
\label{pde} \es
In verifying (\ref{pde}), one finds that the normal-ordering prescription is  
\bs
\be 
: g(\T,\bz,z) \J_a (z) : 
\; =\J_a^- (z) g (\T,\bz,z) + 
g (\T,\bz,z) (\J_a^+ (z) +\J_a (0) \frac{1}{z}) 
\label{cno} 
\ee
\be
: g(\T,\bz,z)  \bJ_a (\bz) : 
\; = \bJ_a^- (\bz) g (\T,\bz,z) + 
g (\T,\bz,z) ( \bJ_a^+ (\bz) + \bJ_a (0) \frac{1}{\bz}) 
\ee
\be
\J_a^{\pm} (z) \equiv  \sum_{m>0}\J_a(\pm m) z^{\mp m -1} 
\sp 
\bJ_a^{\pm} (\bz) \equiv  \sum_{m>0} \bJ_a(\pm m) \bz^{\mp m -1} 
\label{emd} \ee
\label{nop} \es
where the positive and negative modes of the currents are collected in 
the definitions (\ref{emd}) and $J_a(0)$, $\bJ_a (0)$ are the zero modes.
It is easily checked that the PDE's (\ref{cde}) and (\ref{ade}) are 
consistent (that is, (\ref{pde}) defines a flat connection)
because $\J$ commutes with $\bJ$. It will also be convenient to
define the integrated quantities
\be 
Q_a^{\pm} (z) \equiv \pm i \sum_{m>0} {\J_a(\pm m) \over m} z^{\mp m } 
\sp
\bQ_a^{\pm} (\bz) \equiv \pm i \sum_{m>0} {\bJ_a(\pm m) \over m} \bz^{\mp m } 
\label{qdf} \ee
for use below.

Ultimately, one is interested in the $n$-point WZW correlators of the 
WZW primary fields
\be
A_g(\T,\bz, z) = \langle 0 |  g (\T^1,\bz_1,z_1) \cdots g (\T^n,\bz_n,z_n)
| 0 \rangle
\label{wco} \ee
which satisfy the $(g\ti g)$-global Ward identities 
\be
\sum_{i=1}^n \T_a^i A_g(\T,\bz,z)  =  A_g(\T,\bz,z) \sum_{i=1}^n \T_a^i = 0  
\pe 
\label{ggi} \ee
The WZW correlators (\ref{wco}) also satisfy $SL(2,\R) \ti SL(2,\R)$ 
Ward identities
and the chiral and antichiral  Knizhnik-Zamolodchikov (KZ) equations \cit{7}  
\bs
\be
\pa_i A_g(\T,\bz,z)
=A_g(\T,\bz,z)2L^{ab}_g  \sum_{j \neq i} {\T_a^i \T_b^j \over z_{ij} } 
\label{hlk} \ee
\be 
\bpa_i A_g(\T,\bz,z)=
 2L^{ab}_g \sum_{j \neq i} {\T_a^i \T_b^j \over \bz_{ij} } A_g(\T,\bz,z)
\ee
\es
which follow from the PDE's (\ref{pde}) 

\vs .4cm
\ni \un{Semiclassical expansion} 
\vs .3cm

In this paper, we will be interested primarily in the semiclassical or
high-level expansion [44,45,10,43,42] 
 of the low-spin sector of the theory, which is defined
by the level-orders:  
\bs
\be
\J_a(0) = \cO (k^0) \sp 
\J_a(m \neq 0 ) = \cO (k^{1/2}) 
\ee
\be
\bJ_a(0) = \cO (k^0) \sp 
\bJ_a(m \neq 0 ) = \cO (k^{1/2}) 
\ee
\be
\T_a = \cO (k^0)
\ee
\be
L^{ab}_g = {\et^{ab} \over 2k} + \ok = \oko  
\sp 
\D_g (\T) = {Q(\T) \over 2k } +\ok  = \oko \pe  
\ee
\es
In this case, the corresponding semiclassical affine-Sugawara 
conformal blocks and WZW correlators have been worked out in Ref.\cit{42}. 
In particular,
the solution for the high-level $n$-point WZW correlators (\ref{wco}) is 
\be
\label{wzw}
A_g(\T,\bz,z) =  
\left( \one +  2L^{ab}_g
\sum_{i<j}^n  \T_a^i \T_b^j \ln \bz_{ij} \right) I_g^n   
\left( \one +  2 L^{ab}_g \sum_{i<j}^n \T_a^i \T_b^j \ln z_{ij} \right) 
+ \ok 
\ee
$$
 = \left( \one + 4 L^{ab}_g
\sum_{i<j}^n  \T_a^i \T_b^j \ln | z_{ij} |  \right) I_g^n   
+ \ok 
$$ 
where $I_g^n$ is the $n$-point Haar integral
\be 
(I_g^n)_{\a}{}^{\b}
= \int {\rm d} \G \, \G(\T^1)_{\a_1}{}^{\b_1}  \cdots
\G(\T^n)_{\a_n}{}^{\b_n} 
\label{hin} \ee
over unitary Lie group elements $\G(\T)$ in matrix irrep $\T$ of $g$. The
 Haar integral is invariant under $g\ti g$ and satisfies
$(I_g^n)^2 = I_g^n$, so that $I_g^n$ is the projector onto the
$g$-invariant subspace of $\T^1 \oti \cdots \oti \T^n$.  

\vs .4cm
\ni \un{Factorization of the WZW primary fields} 
\vs .3cm 

Our goal in this paper is to solve the algebra (\ref{ega}) and the  PDE
(\ref{pde}) to obtain the explicit semiclassical form (the WZW vertex operators)
of the WZW primary fields $g(\T,\bz,z)$. 
To this end, we look for solutions in the factorized form
\be
g(\T,\bz,z)_\a{}^\b  = \gm (\T,\bz)_\a{}^A \gp (\T,z)_A{}^\b 
\label{fas} \ee
where $A,B = 1 \ldots {\rm dim}\,\T$ are the quantum group indices discussed
by Moore and Resheti-\linebreak khin \cit{25} and others [26-41].  
In what follows, $\gp$ and $\gm$
will be referred to as the chiral and antichiral primary fields and/or the
chiral and antichiral vertex operators of the theory.  

To find such factorized solutions, we assume that $\gp$ and $\gm$ are 
 affine-primary fields under $\J$ and $\bJ$ respectively,
\bs
\be
[\J_a(m),  \gp (\T,z)] = \gp (\T,z) z^m  \T_a 
\label{egp} \ee
\be 
\;\; \; [\bJ_a(m),  \gm (\T,\bz)] = -\bz^m  \T_a \, \gm (\T,\bz) 
\label{egm} \ee
\label{aga}
\es 
which solves (\ref{egc},b) so long as 
\be
[\J_a(m),  \gm (\T,\bz)] \gp (\T,z) = 
\gm (\T,\bz) [\bJ_a(m),  \gp (\T,z)] = 0 \pe  
\label{agb} \ee
Then the PDE's (\ref{cde},b) are solved by the ordinary differential equations
(ODE's)  for the chiral and antichiral primary fields [26,25]  
\bs
\be
\pa \gp(\T,z)  = 2 L^{ab}_g : \gp(\T,z)\J_a(z)  : \T_b 
\label{cod} \ee
\be
\;\;
\bpa \gm(\T,\bz)  = -2 L^{ab}_g \T_b : \gm(\T,\bz) \bJ_a(\bz)  : 
\ee
\label{ode} \es 
where normal ordering is defined by (\ref{nop}) with $g \ra g_{\pm}$.

The explicit semiclassical forms of $g_{\pm}$ (the chiral and antichiral
vertex operators) obtained below by solving the
ODE's (\ref{ode}) will reproduce the 
semiclassical WZW correlators (\ref{wco}) in the factorized form  
\bs
\be
A_g(\T, \bz,z)_\a{}^\b = A_g^- (\T,\bz)_\a{}^A A_g^+ (\T,z)_A{}^\b 
\ee
\be
A_g^+ (\T,z)_A{}^\b 
={}_+\langle 0|  
\gp(\T^1,z_1)_{A_1}{}^{\b_1} \cdots\gp(\T^n,z_n)_{A_n}{}^{\b_n} 
|0 \rangle_+    
\label{cco} \ee
\be
A_g^- (\T,\bz)_\a{}^A 
={}_-\langle 0|  
\gm(\T^1,\bz_1)_{\a_1}{}^{A_1} \cdots\gm(\T^n,\bz_n)_{\a_n}{}^{A_n} 
|0 \rangle_-    
\label{aco} \ee
\be
\sum_{i=1}^n \T_a^i A_g^-(\T,\bz)  =  A_g^+(\T,z) \sum_{i=1}^n \T_a^i = 0  
\label{fgw} \ee
\es
where $A_g^+$ and $A_g^-$ are the chiral and antichiral correlators. 
Here  we have also assumed factorization of the vacuum state
\be
|0 \rangle  =   |0 \rangle_-   |0 \rangle_+    
\ee
into the affine vacua 
$|0 \rangle_+$ and  $ |0 \rangle_-  $ of $\J$ and $\bJ$. We will   
generally suppress the subscripts on the vacua, which will be clear in
context. 

\vs .4cm
\ni \un{The problem}
\vs .3cm

The ODE's (\ref{ode}) for the primary fields $g_{\pm}$
can be solved by iteration of equivalent
integral equations, e.g.
\be
 \gp (\T,z) 
= \gp (\T,z_0) 
+ \int_{z_0}^z \rd z' \; 2 L^{ab}_g  :\gp (\T,z') \J_a(z')  : \T_b  
\label{ieq} \ee
where $z_0$ is a regular reference point. As noted by Moore and Reshetikhin 
\cit{25},
the iterative solution of (\ref{ieq}) is not directly useful because the
leading term $\gp(\T,z) \simeq \gp (\T,z_0)$ of this expansion would give
singular chiral correlators 
$\langle \gp(\T^1,z_0) \cdots  \gp(\T^n,z_0) \rangle = \infty $. In fact,
the iterative solution is somewhat misleading because differentiation of
(\ref{ieq}) by $z_0$ shows that $\gp(\T,z)$ is independent of $z_0$ when the
initial condition $\gp(\T,z_0)$ itself satisfies the original equation
\be
 \pa_{z_0} \gp (\T,z_0 )= 2 L^{ab}_g  : \gp (\T,z_0)\J_a(z_0) : \T_b 
\pe
\ee
Using this fact, we shall see below that the iterative solutions 
can be rearranged into well-defined $z_0$ and $\bz_0$-independent
semiclassical expansions of the primary fields $g_{\pm}$.  

\newpage
\section{Abelian Vertex Operators}

Because it is the simplest, we consider first the case of 
decompactified abelian $g=U(1)^N$, for which the chiral system takes 
the form 
\bs
\be
 \gp (\T,z) 
= \gp (\T,z_0) 
+ \int_{z_0}^z \rd z' \; 2 L^{ab}_g  :\gp (\T,z') \J_a(z')  : \T_b  
\label{aie}  \ee
\be
[\J_a(m),\J_b(n) ] =  k m\et_{ab} \d_{m+n,0} 
\label{aec} \ee
\be
[\J_a(m),  \gp(\T,z)] = \gp(\T,z) z^m   \T_a  
\label{agc} \ee
\be 
[\T_a,\T_b] =0 \sp a,b = 1 \ldots N  
\ee
\be
L_g^{ab} = {\et^{ab} \over 2k} 
\sp
\D_g(\T) ={ \et^{ab} \T_a \T_b \over 2k }  \pe 
\label{hcw} \ee
\label{asy} \es
Here, the representation matrices $\T_a$ (the momenta), and hence
$\gp$, are $1\ti 1$ (i.e. numbers), and we shall see that the solution 
$\gp (\T,z)$ of the system (\ref{asy}) can be rearranged into the
familiar  $z_0$-independent abelian vertex operator of the open bosonic string.
Essentially the same vertex operators (with diagonal matrix $\T_a$'s) are
obtained for compactified abelian algebras such as the Cartan subalgebra of
a Lie algebra or other momentum lattices.

The solution of the integral equation (\ref{aie}) is obtained on inspection as
\be
\label{as1}
\eqalign{
\gp (\T,z) 
= & \exp \left( \int_{z_0}^z \rd z' \, 2 L^{ab}_g \T_a\J_b^-(z')  \right)  
\gp (\T,z_0) \cr  
& 
\ti  \exp \left( \int_{z_0}^z \rd z' \, 2 L^{ab}_g \T_a\J_b(0)/z'   \right)  
 \exp \left( \int_{z_0}^z \rd z' \, 2 L^{ab}_g \T_a\J_b^+(z')   \right)  
 \cr} 
\ee
where $z_0$ is the reference point and $J^{\pm} (z) $ are defined in 
(\ref{emd}).
Performing the integrations in (\ref{as1}), the result may be rearranged 
in the $z_0$-independent form 
\bs
\be
\gp (\T,z) = V_-(\T,z) \Gp(\T) V_0 (\T,z) V_+ (\T,z) 
\label{as2} \ee
\be
V_{\pm} (\T,z) \equiv  \exp(  2 i  L^{ab}_g \T_a  Q_b^{\pm}(z) )
\ee
\be
V_0(\T,z) \equiv  \exp(  2 L^{ab}_g \T_a \J_b(0) \ln z  )
 =z ^{2  L^{ab}_g \T_a \J_b(0) } 
\ee
\es 
where $Q^{\pm}$ is defined in (\ref{qdf}). In this form, we have collected all
$z_0$-dependent factors from the integrations into the constant quantity 
\bs
\be
 \Gp(\T) \equiv V_-(\T,z_0)^{-1} \gp(\T,z_0) V_0(\T,z_0)^{-1} 
V_+ (\T,z_0)^{-1}  
\label{zm1} \ee
\be
\pa_z \Gp = \pa_{z_0} \Gp =0 
\ee
\es  
which is in fact independent of $z_0$ because
 $ \pa_{z_0} \gp (\T,z_0 )= 2 L^{ab}_g  : \gp (\T,z_0)\J_a(z_0) : \T_b $.
In what follows, we refer to the quantity $\Gp (\T) $ as the chiral
zero mode of the chiral vertex operator.

The $z_0$-independent solution (\ref{as2}) has the form of the usual abelian
vertex operator, but we do not yet know the algebra of the zero mode $\Gp(\T)$
 with the currents.

In fact, this algebra is determined by the system. To see this, invert 
(\ref{as2}) to write the zero mode in terms of the primary field  
\be
\Gp(\T) = V_-(\T,z)^{-1} \gp(\T,z) V_+ (\T,z)^{-1} V_0 (\T,z)^{-1} \pe  
\label{zm2} \ee
Then, the algebra of the currents with the zero mode 
\be
[\J_a (m), \Gp (\T) ] =  \Gp (\T)  \T_a \d_{m,0} 
\label{zma} \ee
is obtained straightforwardly from (\ref{zm2}), (\ref{agc}) 
and the current algebra 
(\ref{aec}). 

The algebra (\ref{zma}) is solved by 
\be
\Gp (\T) = {\rm e}^{i q^a \T_a}  
\sp 
[q^a,\J_b(m)] = i \d_b^a \d_{m,0}  
\ee
so that the solution (\ref{as2}) may be written as 
\be
\gp (\T,z) 
= {\rm e}^{ 2 i L_g^{ab} \T_a q_b }  z^{ 2 L^{ab}_g \T_a \J_b(0)}  
\exp(  2 i L^{ab}_g \T_a  Q_b^-(z) ) \exp(  2 i L^{ab}_g \T_a  Q_b^+(z) )
\label{abv} \ee
where $q^a = 2 L_g^{ab} q_b$. With the conventional identification of the
metric $G_{ab}$ and its inverse $G^{ab}$, 
\bs
\be
[\J_a(m),\J_b(n) ] =   m G_{ab} \d_{m+n,0} 
 \ee
\be
G_{ab} = k \et_{ab} \sp G^{ab} = {\et^{ab} \over k } = 2 L_g^{ab} 
\ee
\es
the result (\ref{abv}) is recognized as the familiar abelian
vertex operator with momentum $\T_a$.

As an introduction to the non-abelian case below, we list some well-known
properties of the abelian vertex operators.

\ni {\bf A.} Affine-primary states. On the affine vacuum, the vertex operators
create the affine-primary states
\bs
\be
| \T \rangle \equiv \gp (\T,0) | 0 \rangle = 
\Gp (\T) | 0 \rangle =  
{\rm e}^{ i q^a \T_a } | 0 \rangle 
\ee
\be
\J_a (m \geq 0) | \T \rangle = \d_{m,0} | \T \rangle \T_a
\ee
\es 
which are nothing but the chiral zero modes 
$\Gp (\T)$ on the vacuum. 

\ni {\bf B.} Intrinsic monodromy. When $z$ is taken around a closed loop, one
finds the intrinsic monodromy relation
\be  
\gp (\T,z e^{2 \pi i }) = \gp (\T,z)  e^{4 \pi i L^{ab}_g \T_a\J_b(0) } \pe  
\label{aim} \ee
Using the algebra (\ref{agc}) and the $g$-global invariance (momentum
conservation) in (\ref{ggi}), the operator relation (\ref{aim})
implies the correlator monodromies
\be
\label{imr} 
\eqalign{
\langle 0| & \gp (\T^1,z_1) \cdots
\gp (\T^i,z_i e^{2 \pi i} ) \cdots  \gp (\T^n,z_n) |0 \rangle \cr
 & =\langle 0| \gp (\T^1,z_1 ) \cdots
\gp (\T^i,z_i ) \cdots  \gp (\T^n,z_n) |0 \rangle
{\rm e}^{- 4 \pi i  L^{ab}_g \T_a^i \sum_{j \leq i}^n \T_b^j } \cr }  
\ee
and it is not difficult to see that these phases are trivial for the open
bosonic string.

\ni {\bf C.} Operator products and expansions. 
The operator product of two chiral vertex operators
can be written as  
\bs
\be
\gp (\T^1,z_1) \gp (\T^2,z_2) = \;  
: \gp (\T^1,z_1) \gp (\T^2,z_2) :  z_{12}^{2L^{ab}_g \T_a^1 \T_b^2}   
\;\;\;\;\;\;\;\;\;\;\;\;\; \ee
\be
\label{ano} \eqalign{
: \gp (\T^1,z_1) \gp (\T^2,z_2) : \;  
\equiv  \;  &  \Gp (\T^1) \Gp (\T^2) 
  V_0 (\T^1,z_1) V_0 (\T^2,z_2) \cr  
& \ti  V_-(\T^1,z_1) V_-(\T^2,z_2) 
V_+ (\T^1,z_1) V_+ (\T^2,z_2) \cr}  
\ee
\label{op1} \es 
where $z_{12}=z_1-z_2$ and
the normal-ordered product in (\ref{ano}) puts the
 zero modes $\Gp$ to the left. The closed algebra of the zero modes
\be
\Gp (\T^1) \Gp (\T^2) = \Gp (\T^3) \sp \T^3 \equiv \T^1  +\T^2 
\label{zmr} \ee
then implies the OPE of two vertex operators
\be
\label{op2} 
\eqalign{
\gp  (\T^1,z_1)
& \gp (\T^2,z_2)
  = z_{12}^{\D^g (\T^3) - \D^g(\T^1) -\D^g (\T^2)} 
\left\{  
 \gp (\T^3,z_2 )   \phantom{  { z_{12}^{r+s} \over (r+s+2) !}}  
\right. 
\cr
 & + \sum_{r=0}^{\infty} {z_{12}^{r+1} \over (r+1) !}   
2L^{ab}_g :   \gp (\T^3,z_2)  \pa_2^r\J_b (z_2) : 
\T^1_a \cr  
&  +  \sum_{r,s=0}^{\infty} { z_{12}^{r+s+2} \over (r+s +2) !} 
     4 L^{ab}_g L^{cd}_g
: \gp (\T^3,z_2)
\pa_2^r [ \pa_2^s\J_b(z_2)\J_c(z_2)] :   \T^1_d \T^1_a    \cr  
 & \left.
 \phantom{ \sum_{r=0}^{\infty} {z_1^r \over r !} } 
+ \mbox{higher affine secondaries} \right\} \cr} 
\ee 
where we have used the expression (\ref{hcw}) for the affine-Sugawara conformal
weights. In (\ref{op2}), the normal-ordered product 
$:\gp \J : $ is the chiral analogue of (\ref{cno}), while 
\be
:\gp (\T,z)\J_a(z)\J_b(z) : \; =  \;   
: \,( : \gp (\T,z)\J_b(z) :) \J_a(z) :   
\label{dno}  \ee
is defined iteratively from (\ref{cno}). 
The term ``higher affine secondaries" 
stands for the fields $: \gp \J^p : $, $p \geq 3$
and derivatives thereof.

The OPE (\ref{op2}) follows directly from (\ref{op1}), without
using the ODE (\ref{cod}). Using the ODE however, 
one may rearrange (\ref{op2}) in terms of the affine-primary fields
(and their Virasoro descendants) plus those 
affine-secondary fields which are Virasoro primary (and their Virasoro
descendants). The first few terms of this expansion are  
\be
\label{rop} \eqalign{
\gp (\T^1,z_1)
\gp (\T^2,z_2)
 & = z_{12}^{\D^g (\T^3) - \D^g(\T^1) -\D^g (\T^2)} 
\left\{ 
 \left[ 1 + \frac{1}{2} \sum_{r=1}^{\infty} {z_{12}^r \over r !} \pa_2^r 
\right]   \gp (\T^3,z_2 )    
\right. 
\cr
+  \sum_{r=0}^{\infty} {z_{12}^{r+1} \over (r+1) !} \pa_2^r &  
\left. \{2L^{ab}_g :   \gp (\T^3,z_2) \J_b (z_2) : \} \frac{1}{2} 
[\T^1_a- \T^2_a ]   
 + \ldots \right\} \cr} 
\ee
and the omitted terms are of the form $: \gp \J^p : $, $p \geq 2$
and derivatives thereof. 

\ni {\bf D.} Braid relation. To discuss braiding, we will use the 
Euclidean continuation formula
\be
(z - w)   = (w -z)   e^{ i \pi {\rm sign} ({\rm arg} (z/w)) }
\label{acf} \ee
where $|z| > |w|$ on the left and  
$|w| > |z|$ on the right. The phase in (\ref{acf}) is obtained by requiring
agreement with the corresponding
 computations for vertex operators on the Minkowski world sheet 
(where $z$ and $w$ are on the unit circle).

The braid relation of two chiral vertex operators is then
\bs
\be
\gp (\T^1,z_1) \gp (\T^2,z_2)
  = \B (\T^1\T^2) \gp (\T^2,z_2)  \gp (\T^1,z_1)   
\ee
\be 
\B (\T^1\T^2)  
= e^{ i \pi [\D^g(\T^3) -\D^g(\T^1) - \D^g(\T^2) ] 
{\rm sign} ({\rm arg} (z_1/z_2)) }
\sp 
\B (\T^2\T^1) 
= \B ^{-1} (\T^1\T^2) 
\ee
\label{chb} \es
where $\B$ is the $1 \ti 1$ braid matrix of the abelian theory.

\ni {\bf E.} Chiral correlators. The chiral correlators exhibit the
Koba-Nielsen factor
\be
A_g^+(\T,z)  = 
\langle 0| \Gp (\T^1) \cdots  \Gp (\T^n) |0 \rangle   
 \prod_{i < j}^n z_{ij}^{2L^{ab}_g \T_a^i \T_b^j} 
 = \prod_{i < j} z_{ij}^{\D^g(\T^i +\T^j) - \D^g(\T^i) - \D^g(\T^j) } 
\d( \sum_{i=1}^n \T^i  )  
\label{acc} \ee
where $\d$ is Dirac delta function. 

\ni {\bf F.} Antichiral sector. The antichiral vertex operators are obtained in
the same way,
\bs
\be
\gm (\T,\bz) 
=  \exp(  -2 i L^{ab}_g \T_a  \bQ_b^-(\bz) ) \Gm (\T) 
\bz^{ -2 L^{ab}_g \T_a  \bJ_b(0)}  
\exp( - 2 i L^{ab}_g \T_a  \bQ_b^+(\bz) )
\;\;\;\;\;\;\;\;\;\;\;\;\; 
\ee
\be
\;\;\;\;\;\;\;\;\;\;\;\;\;\;
 = \bz^{-2 \D^g(\T) } \exp(  -2 i L^{ab}_g \T_a  \bQ_b^-(\bz) ) 
\bz^{ -2 L^{ab}_g \T_a  \bJ_b(0)}  
\exp( - 2i L^{ab}_g \T_a  \bQ_b^+(\bz) ) \Gm (\T)  
\label{acv} \ee
\be
\Gm (\T) = {\rm e}^{ -i \bar{q}^a \T_a}  
\sp [  \bar{q}^a , \bJ_b (m)] = i \d_b^a \d_{m,0}  
\ee
\es 
where (\ref{acv}) is written with the antichiral zero mode $\Gm (\T)$ 
on the right. The intrinsic monodromy relations of the antichiral sector are 
\bs
\be
\gm (\T, \bz e^{-2 \pi i }) = 
e^{2 \pi i [ \D_g(\T) + 2 L^{ab}_g \T_a \bJ_b(0) ] }  
\gm (\T,\bz)  
\;\;\;\;\;\;\;\;\;\;\;\;\;\;\;\;\;\; 
\;\;\;\;\;\;\;\;\;\;\;\;\;\;\;\;\;\; 
\ee 
\be
\label{amr} 
\eqalign{
\langle 0| & \gm (\T^1,\bz_1) \cdots
\gm (\T^i,\bz_i e^{-2 \pi i} ) \cdots  \gm (\T^n,\bz_n) |0 \rangle \cr 
 & = {\rm e}^{ 4 \pi i  L^{ab}_g \T_a^i \sum_{j \leq i}^n \T_b^j }  
\langle 0| \gm (\T^1,\bz_1 ) \cdots
\gm (\T^i,\bz_i ) \cdots  \gm (\T^n,\bz_n) |0 \rangle \cr} 
\ee
\es
and we note that the phases in (\ref{amr})
are opposite to those of the chiral sector in (\ref{imr}). The OPE of two
antichiral vertex operators and the antichiral correlators $A_g^-(\T,\bz)$
may be obtained from (\ref{op2}) and (\ref{acc})
with $+ \ra -$,  $z \ra \bz$ and $\T \ra - \T$.

\ni {\bf G.} Nonchiral results. 
Combining the chiral and antichiral vertex operators,
we have the nonchiral vertex operators
\bs
\be 
g(\T,\bz,z) = \gm (\T,\bz) \gp (\T,z)
\;\;\;\;\;\;\;\;\;\;\;\;\;\;\;\;\;\;\;\;\;\;\;\;\;\;\;\;\;\;\;\;\;\;\;\;\;\;\;  
\phantom{\exp(  2 L^{ab}_g \T_a  \bQ_b^+(\bz) ) G (\T)}  
\ee
\be
\eqalign{\;\;\;\;\;\;\;\;\;\;\;\;\;\; =\bz^{-2 \D^g(\T) }  
   & \exp(  -2 i L^{ab}_g \T_a  \bQ_b^-(\bz) ) 
\bz^{ -2 L^{ab}_g \T_a  \bJ_b(0)}  
\exp(  -2 i L^{ab}_g \T_a  \bQ_b^+(\bz) ) G (\T) \cr  
    & \ti \exp(  2 i L^{ab}_g \T_a  Q_b^-(z) )
z^{ 2 L^{ab}_g \T_a \J_b(0)}  
\exp(  2 i L^{ab}_g \T_a  Q_b^+(z) ) \cr }
\ee
\be
G(\T) \equiv  \Gm (\T)  \Gp (\T)
={\rm e}^{  i (q^a - \bq^a) \T_a } 
\ee
\label{anv} \es
where we have made the conventional assumption (which solves (\ref{agb}))
that $q_a$ ($\bq_a$)  commutes with all the operators of the antichiral
(chiral) sector. This assumption is examined in further detail for the
nonabelian case in Section 7. 

The nonchiral vertex operators (\ref{anv}) give the OPE's and 
nonchiral correlators     
\bs
\be 
\eqalign{
\;\;\;\;\;\;\;\; g & (\T^1,\bz_1,z_1)
g (\T^2,\bz_2,z_2) \cr 
& = |z_{12}|^{2[\D^g(\T^1 +\T^2) -\D^g(\T^1) -\D^g(\T^2)]}  
\left\{  
 g (\T^3,\bz_2,z_2 )   \phantom{  { z_{12}^{r+s} \over (r+s+2) !}}  
\right. 
\cr
& + \sum_{r=0}^{\infty} {z_{12}^{r+1} \over (r+1) !}   
2L^{ab}_g :   g (\T^3,\bz_2,z_2)  \pa_2^r\J_b (z_2) : 
\T^1_a \cr  
& -  \sum_{r=0}^{\infty} {\bz_{12}^{r+1} \over (r+1) !}   
2L^{ab}_g :   g (\T^3,\bz_2,z_2)  \bpa_2^r \bJ_b (\bz_2) : 
\T^1_a \cr  
 & +  \sum_{r,s=0}^{\infty} { z_{12}^{r+s+2} \over (r+s+2) !} 
      4 L^{ab}_g L^{cd}_g : g (\T^3,\bz_2,z_2)
\pa_2^r [ \pa_2^s\J_b(z_2)\J_c(z_2)] :   \T^1_d \T^1_a    \cr  
 & +  \sum_{r,s=0}^{\infty} { \bz_{12}^{r+s+2} \over (r+s+2) !} 
      4 L^{ab}_g L^{cd}_g : g (\T^3,\bz_2,z_2)
\bpa_2^r [ \bpa_2^s \bJ_b(\bz_2) \bJ_c(\bz_2)] :  \T^1_a \T^1_d    \cr  
& - \sum_{r,s=0}^{\infty} {z_{12}^{r+1} \over (r+1) !} 
{\bz_{12}^{s+1} \over (s+1) !} 
 4 L^{ab}_g L^{cd}_g :   
g (\T^3,\bz_2,z_2)  \pa_2^r\J_b (z_2) \bpa_2^s \bJ_c (\bz_2)  : 
\T^1_d \T^1_a     \cr  
& 
\left.
\phantom{ \sum_{r=0}^{\infty} {z_{2}^r \over r !} }   
 + \mbox{higher affine secondaries} \right\} \cr} 
\ee 
\be  
A_g (\T,\bz,z) = A_g^- (\T,\bz) A_g^+ (\T,z) 
= \prod_{i < j} |z_{ij}|^ {2[\D^g(\T^i +\T^j) -\D^g(\T^i) -\D^g(\T^j)]} 
\d^2( \sum_{i=1}^n \T^i   ) \pe  
\label{anc} \ee
\es
 Both of these results show trivial monodromy when any $z$ is taken around 
another, and, moreover,  
the Virasoro-Shapiro factor in (\ref{anc}) shows that the intrinsic
monodromies (\ref{imr}) and (\ref{amr}) have cancelled as they should.
  
\newpage
\section{Semiclassical Nonabelian Vertex Operators 
$\;\;\;\;\;\;\;\;\;\;\;\;$  
for the Affine-Sugawara Constructions}

\ni \un{Chiral sector}
\vs .3cm

The defining relations for the general
chiral fields $\gp (\T,z)$ are
\bs
\be
 \gp (\T,z) 
= \gp (\T,z_0) 
+ \int_{z_0}^z \rd z' \; 2 L^{ab}_g  :\gp (\T,z') \J_a(z')  : \T_b  
 \label{ien} \ee
\be
[\J_a(m),\J_b(n) ] = if_{ab}{}^c\J_c(m+n) +  k m\et_{ab} \d_{m+n,0} 
\label{gec} \ee
\be
[\J_a(m),  \gp(\T,z)] = \gp(\T,z) z^m   \T_a  
\sp [\T_a,\T_b] = i f_{ab}{}^c  \T_c  
\label{ggc} \ee
\be 
\J_a (m \neq 0) = \okh \sp 
\J_a (0) = \cO (k^0) 
\label{lor} \ee
\be
\gp (\T,z)  = \cO (k^0) 
\sp \T_a = \cO (k^0) 
\ee
\be
 L^{ab}_g = \oko
\sp \D_g(\T) = \oko
\ee
\es
where $z_0$ is a regular reference point and $L_g^{ab}$ is given in (\ref{asl}).
Using the results of the previous section as a guide, and paying close 
attention to the level-orders (\ref{lor}-f),  the iterative
solution of (\ref{ien}) can be rearranged into a $z_0$-independent 
semiclassical or high-level expansion of $\gp$.

We give here the solution of (\ref{ien}) up to  $\okt $:
\be 
\gp (\T,z)  = \pG (\T) 
+  
2 i L^{ab}_g [ Q_a^- (z) \pG (\T) + \Gp (\T)  Q_a^+ (z)
- \pG (\T) i \J_a(0) \ln z ] \T_b
\label{vop} \ee
$$
+ 4 L^{ab}_g L^{cd}_g \left\{  
\sum_{m,n >0}  [\J_b(-m) {\J_c(-n)\over n}\pG(\T){z^{m+n} \over m+n }   
+  \pG(\T ){\J_c(n) \over n}\J_b(m)  {z^{-(m+n)} \over m+n }  ]  
\right. 
$$ $$
- \sum_{m,n >0 \atop m \neq n} 
 [\J_b(-m) \pG (\T)  {\J_c(n) \over n}  {z^{m-n} \over m-n }   
+   {\J_c(-n) \over n} \pG (\T)\J_b(m)  {z^{-(m-n)} \over m-n }  ] 
$$ $$
\left. 
+ \sum_{n >0 } 
  [ {\J_c(-n) \over n} \pG(\T) \J_b(n) (\ln z  - \frac{1}{2n} ) 
-\J_{b}(-n) \pG(\T) {\J_c(n) \over n}(\ln z + \frac{1}{2n} )]\right\} \T_d \T_a 
 $$
$$
+\okt \pe  
$$
This is the explicit semiclassical form of the new nonabelian chiral vertex
operators. Here, $\Gp (\T) $ is the constant chiral zero mode, which carries
the index structure $\Gp(\T)_A{}^\a$ (in parallel with $\gp$), and
which satisfies
\bs
\be
\Gp (\T) = \cO (k^0)  
\label{ocz} \ee
\be
\pa \Gp (\T) = 0+  \okt \pe 
\label{dcz} \ee
\es
Using the level-orders (\ref{lor}-f) and (\ref{ocz}), one sees that the 
terms proportional to $Q_{\pm}$ in (\ref{vop}) 
are $\cO (k^{-1/2})$, while the rest of the explicit terms are $\oko$.
Using (\ref{dcz}), it is straightforward to check by differentiation that the
chiral vertex operator (\ref{vop}) satisfies
\be
\pa \gp (\T,z) = 2 L^{ab}_g  : \gp (\T,z)\J_a (z)    : \T_b   + \okt  
\label{odc} \ee
as it should.

The result (\ref{vop})  can be inverted to write the zero mode 
$\Gp$ in terms of the primary field $\gp$,  
\be
\pG (\T) = \gp(\T,z)  - 2i L^{ab}_g [ 
 Q_a^- (z)\gp(\T,z)  
 + \gp (\T,z)  Q_a^+ (z) 
-\gp (\T,z) i \J_a(0) \ln z ] \T_b
\label{czm} \ee
$$
+ 4 L^{ab}_g L^{cd}_g \left\{  
\sum_{m,n >0} [{\J_b(-m) \over m}\J_c(-n) \gp (\T,z)  {z^{m+n} \over m+n }   
+  \gp (\T,z) \J_c(n) {\J_b(m) \over m}  {z^{-(m+n)} \over m+n }  ]  
\right. 
$$
$$
+ \sum_{m,n >0 \atop m \neq n} 
[ {\J_b(-m) \over m} \gp (\T,z) \J_c(n)   {z^{m-n} \over m-n }   
+  \J_c(-n) \gp (\T,z) {\J_b(m) \over m}  {z^{-(m-n)} \over m-n }  ]  
$$
$$
\left. 
+ \sum_{n >0 } 
[\J_{b}(-n) \gp (\T,z) {\J_c(n) \over n} (\ln z - {1  \over 2 n} )    
-   {\J_c(-n) \over n} \gp (\T,z) \J_b(n) (\ln z + {1  \over 2 n} )  ]  
\right\} \T_d \T_a  
$$
$$ +\okt   $$
and it is straightforward to check by differentiation with (\ref{odc})
that $\Gp (\T)$ in  (\ref{czm}) satisfies (\ref{dcz}).  
Moreover, by setting $z=z_0$ in (\ref{vop}) and (\ref{czm}) one can see the
intermediate relations between the zero mode $\Gp (\T)$ and the primary field
$\gp (\T,z_0) $  at
the reference point $z_0$ (These relations are the nonabelian analogues of 
(\ref{zm1}) 
and (\ref{zm2})). The zero mode is of course independent of $z_0$
\be
\pa_{z_0} \Gp (\T) = 0 + \okt 
\ee
just as it is independent of $z$, because the differential equation (\ref{odc})
holds as well at the reference point.  

Following the previous section, the next step is to use the inversion
 (\ref{czm}) and the algebra (\ref{ggc}) of the
currents with the primary field $\gp$ to obtain the algebra of the currents 
with the zero mode $\Gp $.  After some algebra, the result is
\bs
\be
[\J_a (0) , \pG (\T) 
] = \pG (\T)   \T_a + \okt 
\label{zmc} \ee
\be
[\J_a (m \neq 0 ) , \pG (\T) ] =  {i \over 2 k m } 
 : \pG (\T)\J_b(m) :  f_{a}{}^{bc}\T_c         
+ \oko  
\ee
\label{gza} \es
which reduces to the algebra (\ref{zma}) 
in the abelian case. It is likely that the relation (\ref{zmc}) is
exact to all orders. 

The algebra (\ref{gza}) shows  that
\be
[\J_a (m \neq 0) , \Gp (\T) ] =   \cO (k^{-1/2} )
\ee
so, without loss of accuracy, we may move any factor 
$\Gp$ in (\ref{vop}) thru any non-zero
moded current. In particular, the chiral vertex operator may be written
with the chiral zero mode on the left
\be
\gp (\T,z) = 
  \Gp (\T)  [ \one +  i X^a(z)  \T_a + N^{ab}(z) \T_b \T_a ]  + \okt  
\label{cvo} \ee
where we have defined the quantities  
\bs
\be
X^a(z) \equiv  2 L^{ab}_g [  
Q_b^-(z)  +  Q_b^+(z) -i  \J_b(0)  \ln z ] 
\label{xde} \ee
\be
\label{nde} 
\eqalign{
N^{ab} (z) 
\equiv  4 L^{ac}_g L^{db}_g & \left\{  
 \sum_{m,n >0} [\J_c(-m) {\J_d(-n) \over n}  {z^{m+n} \over m+n }   
+   {\J_d(n) \over n}\J_c(m)  {z^{-(m+n)} \over m+n }  ]  
\right. \cr  
& -   \sum_{m,n >0 \atop m \neq n} 
[\J_c(-m)  {\J_d(n) \over n}  {z^{m-n} \over m-n }   
+   {\J_d(-n) \over n} \J_c(m)  {z^{-(m-n)} \over m-n }  ]  
\cr 
& \left. +  \sum_{n >0 } 
[ 
  {\J_d(-n) \over n} \J_c(n) (\ln z - \frac{1}{2n} )  
-\J_{c}(-n)  {\J_d(n) \over n} (\ln z + \frac{1}{2n} )    
]  \right\} \pe \cr }  
\ee
\es 
It will also be convenient to have the inversion of (\ref{cvo}) 
\be
\Gp (\T) = 
 \gp (\T,z) [ \one -  i X^a(z)  \T_a -  (N^{ab}(z) + X^b(z) X^a(z))\T_b \T_a ] 
 + \okt  
\label{icv} \ee
which agrees with eq.(\ref{czm}) thru the indicated order.

Restoring the Lie algebra and quantum group indices $\a$ and $A$,
\bs
\be
\gp (\T,z)_A{}^\a  = 
  \Gp (\T)_A{}^\b   [ \one + i X^a(z)  \T_a + N^{ab}(z) \T_b \T_a ]_\b{}^\a
  + \okt  
\ee
\be
\a, A = 1 \ldots {\rm dim}\,\T
\ee
\es
we see  that, thru this order of the semiclassical expansion, the
quantum group acts only on the chiral zero mode $\Gp$. 

We also remark that the algebra (\ref{gza}) is consistent with 
unitarity${}^{\rm a}$\footnotetext{${}^{\rm a}$Unitarity is easiest to check 
on the unit circle (where $z^* =z^{-1}$) using a Cartesian frame with 
 $\et_{ab} =\d_{ab}$, 
$\J_a^\dagger (m) =\J_a (-m)$ and $\T_a$ Hermitean.} of the chiral
zero mode
\be
\Gp^\dagger (\T) \Gp (\T) = \Gp (\T) \Gp^\dagger (\T) = \one + \oko
\ee
which implies that the extreme semiclassical chiral vertex operator
\be
\gp (\T,z) = \Gp (\T) \exp [  2i L^{ab}_g (Q_a^-(z)  + Q_a^+(z) ) \T_b ]  
+ \oko
\ee
is also unitary 
\be
\gp^\dagger (\T,z) \gp (\T,z)= \gp (\T,z) \gp^\dagger (\T,z) = \one + \oko
\ee
thru the indicated order. It is known from the abelian case that vertex 
operators cannot be unitary operators
beyond this order, due to normal ordering, which
enters in $\gp$ (and $\Gp$) at order $k^{-1}$. 

\vs .4cm
\ni \un{Antichiral sector}
\vs .3cm

Following similar steps, we have solved the antichiral ODE
\be 
\bpa \gm (\T,\bz) = -2 L^{ab}_g \T_b : \gm (\T,\bz) \bJ_a (\bz)    :  + \okt 
\label{oda} \ee
for the antichiral vertex operator $\gm (\T,\bz)$ thru the same order. 
The main results are as follows:

\ni {\bf 1.} Antichiral vertex operator and zero mode.
The antichiral vertex operator is
\be 
\gm (\T,\bz) = \bG (\T) - 2 i L^{ab}_g \T_b [ \bQ_a^-(\bz)  \bG (\T) +  
 \bG (\T) \bQ_a^+ (\bz)  - \bG (\T) i \bJ_a(0) \ln \bz ]
\label{avp}
\ee
$$
+ 4 L^{ab}_g L^{cd}_g \T_a \T_d \left\{  
\sum_{m,n >0} [ \bJ_b(-m) {\bJ_c(-n) \over n} \bG (\T) {\bz^{m+n} \over m+n }   
+  \bG (\T) {\bJ_c(n) \over n} \bJ_b(m)  {\bz^{-(m+n)} \over m+n }  ]  
\right. 
$$
$$
- \sum_{m,n >0 \atop m \neq n} 
[ \bJ_b(-m) \bG (\T) {\bJ_c(n) \over n}  {\bz^{m-n} \over m-n }   
+   {\bJ_c(-n) \over n} \bG (\T) \bJ_b(m)  {\bz^{-(m-n)} \over m-n }  ]  
$$
$$
\left. + \sum_{n >0 } 
[ {\bJ_c(-n) \over n} \bG (\T)  \bJ_b(n) (\ln \bz - \frac{1}{2n} )  
-\bJ_{b}(-n) \bG (\T) {\bJ_c(n) \over n} (\ln \bz + \frac{1}{2 n} )    
]  
\right\} 
$$
$$
+\okt 
 $$
where $\Gm$, with index structure $\Gm (\T)_\a{}^A $, 
is the antichiral zero mode:
\bs
\be 
\Gm (\T) = \cO (k^0) 
\ee
\be  
\bpa \Gm (\T) = 0 + \okt  
\pe
\label{zmd}  \ee
\es
Inversion of (\ref{avp}) gives the antichiral zero mode $\Gm$
in terms of
the antichiral primary field $\gm$ 
\be
\bG (\T) = \gm (\T,\bz) + 2 i L^{ab}_g \T_b [ 
 \bQ_a^-(\bz)  \gm (\T,\bz)  
 + \gm(\T,\bz)  \bQ_a^+ (\bz)  
-\gm (\T,\bz)i \bJ_a(0) \ln \bz ] 
\label{azm} \ee
$$
+ 4 L^{ab}_g L^{cd}_g  \T_a \T_d  \left\{  
\sum_{m,n >0}[{\bJ_b(-m) \over m}\bJ_c(-n)\gm (\T,\bz) {\bz^{m+n} \over m+n }   
+  \gm (\T,\bz) \bJ_c(n) { \bJ_b(m) \over m}  {\bz^{-(m+n)} \over m+n }  ]  
\right. 
$$
$$
+ \sum_{m,n >0 \atop m \neq n} 
[ {\bJ_b(-m) \over m} \gm (\T,\bz) \bJ_c(n)   {\bz^{m-n} \over m-n }   
+   \bJ_c(-n) \gm (\T,\bz) { \bJ_b(m) \over m}  {\bz^{-(m-n)} \over m-n }  ]  
$$
$$
\left. + \sum_{n >0 } 
[ \bJ_{b}(-n) \gm (\T,\bz) {\bJ_c(n) \over n} (\ln \bz - {1 \over 2 n} )    
-   {\bJ_c(-n) \over n} \gm (\T,\bz) \bJ_b(n) (\ln \bz + {1  \over 2 n} )  ]  
\right\} $$
$$
 +\okt  $$
and it is not difficult to check that the results (\ref{avp}) and (\ref{azm})
satisfy the differential equations (\ref{oda}) and (\ref{zmd}).

\ni {\bf 2.} Algebra of the zero modes. Using (\ref{azm}) and 
the algebra (\ref{egm}) of the antichiral
currents with the primary field $\gm$, we obtain
the algebra of the antichiral currents with the antichiral zero mode $\Gm$, 
\bs
\be
[ \bJ_a (0) , \bG (\T) ] = - \T_a \bG(\T)  + \okt 
\label{cza} \ee
\be 
[ \bJ_a (m \neq 0 ) , \bG (\T)] =  -{i  \over 2k m}  f_a{}^{bc} \T_c 
: \bG (\T) \bJ_b(m) :  
+ \oko  
\ee
\label{aza} 
\es
so that $[ \bJ_a (m \neq 0 ) , \bG (\T) ] =  \cO (k^{-1/2} ) $ as in the 
chiral sector.

\ni {\bf 3.} Rearrangements.  
Using the algebra (\ref{aza}), a number of alternate forms may be obtained 
for the antichiral vertex operator, 
\bs
\be
\label{ace} 
\gm(\T,\bz) = \bG (\T) -  i \T_a  \bG (\T) \bX^a (\bz) + 
\T_a \T_b   \bG (\T) \bN^{ab} (\bz) +\okt \;\;\;\;\; \;\;\;\;  
\ee
\be
\;\;\;\;\;\;\;\;\; = [ \one   -  i \T_a   \bX^a (\bz)  -2 \D^g (\T) \ln \bz +
\T_a \T_b \bN^{ab} (\bz) ]  \bG (\T) + \okt 
\label{avb}  \ee
\be
 = \bz^{-2 \D^g (\T)} [ \one   -  i \T_a   \bX^a (\bz) +
\T_a \T_b \bN^{ab} (\bz) ]  \bG (\T) + \okt   
\label{avo} 
 \ee 
\be
 \bX^a (\bz) \equiv  X^a(z) \ve_{z \ra \bz, \J \ra \bJ} 
\sp
\bN^{ab} (\bz)  \equiv  N^{ab} (z) \ve_{z \ra \bz, \J \ra \bJ} \pe  
\ee
\label{avt} \es
where $X$ and $N$ are defined in (\ref{xde},b). 
It will also be useful to have the inverse of (\ref{avo}), 
\be
\Gm(\T)    
  = \bz^{2 \D^g (\T)} [ \one   +  i \T_a   \bX^a (\bz)-  
\T_a \T_b (\bN^{ab} (\bz) + \bX^a (\bz) \bX^b(\bz) )  ]  \gm (\T,\bz) + \okt 
 \label{iav} \ee 
which agrees with (\ref{azm}) thru the indicated order.

In (\ref{avb},c) we have chosen to write the antichiral zero mode $\Gm (\T)$ 
on the right. The index structure of $\Gm$ is 
$(\Gm)_\a{}^A$, so these forms show that the 
 quantum group acts only on $\Gm$ thru this  order of the semiclassical
expansion. 
 
\ni {\bf 4.} Semiclassical unitarity.
As in the chiral sector,  the algebra (\ref{aza}) is consistent with 
semiclassical  unitarity of the antichiral
zero mode
\be
\Gm^\dagger (\T) \Gm (\T) = \Gm (\T) \Gm^\dagger (\T) = \one + \oko
\ee
and this implies that the extreme semiclassical antichiral vertex operator
\be
\gm (\T,\bz) = 
 \exp [   -2 iL^{ab}_g ( \bQ_a^- (\bz) + \bQ_a^+ (\bz)  )  \T_b ]  
\Gm (\T) + \oko
\ee
is also unitary 
\be
\gm^\dagger (\T,\bz) \gm (\T,\bz)= \gm (\T,\bz) \gm^\dagger (\T,\bz) = 
\one + \oko
\ee
thru the indicated order. 

\vs .4cm
\newpage
\ni \un{Some simple chiral and antichiral applications}
\vs .3cm

We conclude this section with some simple applications of these results.

\ni {\bf A.} Affine primary fields. By construction, the chiral and antichiral
vertex operators (\ref{cvo}) and (\ref{avo}) are (${\rm dim}\,\T$ 
quantum group ``copies'' of) 
affine-primary fields under their respective affine algebras,
\be
\left. 
\matrix{
[\J_a (m), \gp (\T,z)_A{}^\a ] =   \gp (\T,z)_A{}^\b z^m (\T_a)_\b{}^\a \cr
[\bJ_a (m), \gm (\T,\bz)_\a{}^A ] =  - \bz^m(\T_a)_\a{}^\b \gm (\T,\bz)_\b{}^A
\cr} \right\} 
+ \left\{
\matrix{ 
\okt & {\rm when}\;\, m=0 \cr
\oko & {\rm when}\;\, m\neq 0 \cr} \right.  \pe  
\ee
These relations can also be checked directly using the current algebra and
the algebra (\ref{gza}), (\ref{aza})
 of the currents with the zero modes $G_{\pm}$. On the
affine vacuum, the vertex operators create (copies of) affine-primary states
\bs
\be
\gp (\T,0)_A{}^\a | 0 \rangle = 
\Gp (\T)_A{}^\a | 0 \rangle  + \okt  
\ee
\be
\gm (\T,0)_\a{}^A | 0 \rangle = 
\bG (\T)_\a{}^A | 0 \rangle  + \okt  
\ee
\es
which, as in the abelian case,
 are proportional to the chiral and antichiral zero modes.

\ni {\bf B.} Affine-Sugawara primary fields. 
We have also checked explicitly that
\bs
\be
T_g (z) \gp (\T,w) = \left[ {\D_g (\T) \over (z-w)^2 } + {1 \over z-w} \pa_w
\right] \gp (\T,w) + \okt
\ee
\be
\bT_g (\bz) \gm (\T,\bw) = \left[ {\D_g (\T) \over (\bz-\bw)^2 } + 
{1 \over \bz-\bw} \pa_w
\right] \gm (\T,\bw) + \okt
\ee
\es   
so that, as they should be,
 the chiral and antichiral vertex operators are Virasoro
primary fields under the affine-Sugawara 
constructions $T_g$ and $\bT_g$ respectively.

\ni {\bf C.} Intrinsic monodromies.
 When $z$ is taken around a closed loop, one finds the intrinsic  
monodromy relations,
\bs
\be  
\gp (\T,z e^{2 \pi i }) = \gp (\T,z) \exp \left(  
  4 \pi i [ L^{ab}_g \T_a\J_b(0) + 2 i L_g^{ac} L_g^{db}   
f_{ab}{}^e \T_e \sum_{n >0} {\J_c(-n)\J_d(n) \over n} ] \right) 
\ee
$$
+ \okt  
$$
\be  
\gm (\T,\bz e^{-2 \pi i }) = 
\exp \left( 4 \pi i [  \D^g(\T) + L^{ab}_g \T_a \bJ_b(0) + 
2 i L_g^{ac} L_g^{db}   
f_{ab}{}^e \T_e \sum_{n >0} {\bJ_c(-n) \bJ_d(n) \over n} ] \right)  
\ee
$$
\ti \gm (\T,\bz) 
 + \okt \pe  
$$
\label{gmr} \es 
Using (\ref{ava}), (\ref{gza}), (\ref{aza}) and the
$(g \ti g)$-global Ward identities, the operator
relations (\ref{gmr}) imply the much simpler 
monodromies for the correlators
\bs
\be
\eqalign{
\langle 0| & \gp (\T^1,z_1) \cdots
\gp (\T^i,z_i e^{2 \pi i} ) \cdots  \gp (\T^n,z_n) |0 \rangle \cr 
& =\langle 0| \gp (\T^1,z_1 ) \cdots
\gp (\T^i,z_i ) \cdots  \gp (\T^n,z_n) |0 \rangle
e^{- 4 \pi i  L^{ab}_g \T_a^i \sum_{j \leq i}^n \T_b^j } + \okt \cr }
\ee
\be
\eqalign{
\langle 0| & \gm (\T^1,\bz_1) \cdots
\gm (\T^i,\bz_i e^{-2 \pi i} ) \cdots  \gm (\T^n,\bz_n) |0 \rangle \cr 
 & = e^{ 4 \pi i  L^{ab}_g \T_a^i \sum_{j \leq i}^n \T_b^j }  
\langle 0| \gm (\T^1,\bz_1 ) \cdots
\gm (\T^i,\bz_i ) \cdots  \gm (\T^n,\bz_n) |0 \rangle +\okt \cr} 
\ee 
\label{imn} \es
because the normal-ordered terms in (\ref{gmr}) do not contribute to the
correlators at the indicated order.

\section{Operator Products and Expansions}

In this section, we combine our results above with those of Ref.\cit{25} to
obtain the chiral-chiral and antichiral-antichiral
 semiclassical operator products and OPE's 
of the semiclassical vertex operators (\ref{cvo}) and (\ref{avo}).
The corresponding chiral-antichiral products and expansions are discussed 
in Section 7. 

\vs .4cm
\ni \un{Chiral sector}
\vs .3cm

The product of two chiral vertex operators (\ref{cvo}) can be 
written as
\be
 \gp (\T^1,z_1) \gp (\T^2,z_2) =  \;  
 : \gp (\T^1,z_1) \gp (\T^2,z_2) : (\one +    
 2 L^{ab}_g \T_a^1 \T_b^2 \ln z_{12} )  
 + \okt  
\ee
$$
 \;\;\;\;\;\;\; = : \gp (\T^1,z_1) \gp (\T^2,z_2) : 
 z_{12}^{2 L^{ab}_g \T_a^1 \T_b^2 }   
 + \okt  
$$
where the normal-ordered product of two vertex operators is defined in
parallel to (\ref{ano}),   
\bs
\be
\label{ggo}
\eqalign{
: \gp (\T^1,z_1) \gp (\T^2,z_2) : 
\; \equiv \; &  \pG (\T^1 )  \pG (\T^2 ) 
 [ \one + i X^a(z_1)  \T^1_a + i X^a (z_2)  \T^2_a \cr 
+ &  N^{ab}(z_1) \T^1_b \T^1_a   
+ N^{ab}(z_2) \T^2_b \T^2_a   
+ N_2^{ab} (z_1,z_2) \T_a^1 \T_b^2 ] + \okt \cr} 
 \ee  
\be
\eqalign{
N_2^{ab} (z_1,z_2) 
 & \equiv  \; - : X^a (z_1) X^b(z_2) : \cr 
 & =  \! -4 L^{ac}_g L^{db}_g
[ Q_c^-(z_1) ( Q_d^- (z_2) +  Q_d^+(z_2) ) 
 +  (Q_d^- (z_2) + Q_d^+(z_2) ) Q_c^+ (z_1)  ] + \okt  \cr } 
\ee
\es
with zero modes on the left. 

For high-level
closure of the affine-primary fields, the product
of two zero modes must close into zero modes,
\be 
\Gp(\T^1)_{A_1}{}^{\a_1} 
\Gp(\T^2)_{A_2}{}^{\a_2} 
=\sum_{\T_k, A_k , \a_k }  
F_{A_1 A_2 \a_k}^{\a_1 \a_2 A_k} (\T^1 \T^2 \T^k)  
	\Gp(\T^k)_{A_k}{}^{\a_k}   + \okt  
\label{ggr} \ee 
which is the non-abelian analogue of (\ref{zmr}). Following Ref.\cit{25}, 
we will assume that the fusion coefficient $F$ in (\ref{ggr}) has the
factorized form
\be
F_{A_1 A_2 \a_k}^{\a_1 \a_2 A_k} (\T^1 \T^2 \T^k) 
= Q_{A_1 A_2}{}^{A_k} (\T^1 \T^2 \T^k) 
C^{\a_1 \a_2 }{}_{\a_k}  (\T^1 \T^2 \T^k) + \okt 
\label{fff} \ee
Here $ C  $ is the  
usual Clebsch-Gordon coefficient for $\T^1 \oti \T^2$ into
$\T^k$, which satisfies the  $g$-global Ward identity 
\be
C^{\b_1 \b_2} {}_{\a_k}(\T^1 \T^2 \T^k) [ 
(\T_a^1)_{\b_1}{}^{\a_1} \d_{\b_2}{}^{\a_2} 
+\d_{\b_1}{}^{\a_1} (\T_a^2)_{\b_2}{}^{\a_2}] 
= (\T_a^k)_{\a_k}{}^{\b_k}  
C^{\a_1 \a_2} {}_{\b_k}(\T^1 \T^2 \T^k) 
\label{cgi} \ee
while $Q$ is the corresponding (level-dependent) quantum Clebsch-Gordan 
coefficient.  

Using (\ref{ggr}) and (\ref{cgi}), we may expand the operator product 
(\ref{ggo}) for $z_1$ near $z_2$. After some algebra, one
finds the OPE of two chiral vertex operators 
\be
\label{cop} 
\eqalign{ 
\gp & (\T^1,z_1)_{A_1}{}^{\a_1}   
  \gp (\T^2,z_2)_{A_2}{}^{\a_2} \cr  
 & = \sum_{ \T^k , A_k, \a_k \atop \b_1}  
 z_{12}^{\D^g (\T^k) - \D^g(\T^1) -\D^g (\T^2)} 
F_{A_1 A_2 \a_k}^{\b_1 \a_2 A_k} (\T^1 \T^2 \T^k)  
\left\{ 
 \gp (\T^k,z_2 )_{A_k}{}^{\a_k}  \d_{\b_1}^{\a_1}
 \phantom{  { z_{2}^{r} \over s !}}  
\right. 
\cr
& + \sum_{r=0}^{\infty} {z_{12}^{r+1} \over (r+1) !}   
2L^{ab}_g :   \gp (\T^k,z_2)_{A_k}{}^{\a_k}    \pa_2^r\J_b (z_2) : 
(\T^1_a)_{\b_1}{}^{\a_1}   \cr  
 & +  \sum_{r,s=0}^{\infty} { z_{12}^{r+s+2} \over (r+s+2) !} 
   \left.    4 L^{ab}_g L^{cd}_g 
: \gp (\T^k,z_2)_{A_k}{}^{\a_k}  
\pa_2^r [ \pa_2^s\J_b(z_2)\J_c(z_2)] :   
(\T^1_d \T^1_a)_{\b_1}{}^{\a_1}  \right\}  \cr    
& + \okt \cr}
\ee 
where the normal-ordered product $:\gp \J:$ is given in 
 (\ref{cno}), and 
\be
\label{noe}
\eqalign{
:\gp (\T,z)\J_a(z)\J_b(z) : \;  \equiv \; & 
\J_a^-(z) [\J_b^-(z) \gp (\T,z) + \gp (\T,z)\J_b^+(z)] \cr  
 &  + [\J_b^-(z) \gp (\T,z) + \gp (\T,z)\J_b^+(z)] \J_a^+(z) 
 \pe \cr}  
\ee
The definition (\ref{noe}) can be replaced with (\ref{dno}) to the order
we are working,
since the extra current zero-mode contributions in (\ref{dno}) would
contribute to (\ref{cop}) at $\okt$. 

The right side of the OPE (\ref{cop}) shows the affine-primary fields $\gp$
and an infinite number of affine-secondary fields of the form
$: \gp \J:$, $:\gp \J \J $: and derivatives thereof. One also notes that the  
chiral OPE has the schematic form \cit{42}
\be
\eqalign{
\mbox{affine-primary} \cdot \mbox{affine-primary} 
= & \cO (k^0) \cdot \mbox{affine primaries} \cr  
 &+ \oko \cdot \mbox{affine secondaries} \cr}  
\ee
so that the affine-primary fields close into themselves in the extreme classical
limit.

Using also the high-level ODE (\ref{odc}), one may rearrange (\ref{cop})
in the form  analogous to eq.(\ref{rop})
\be
\gp (\T^1,z_1)_{A_1}{}^{\a_1}   
 \gp (\T^2,z_2)_{A_2}{}^{\a_2}
\phantom{ \C_{A_1 A_2 \a_k}^{\b_1 \b_2 A_k} (\T^1 \T^2 \T^k)  
 z_{12}^{\D^g (\T^k) - \D^g(\T^1) -\D^g (\T^2)}  
 z_{12}^{\D^g (\T^k) - \D^g(\T^1) -\D^g (\T^2)} } 
\label{sop} \ee 
$$
 = \sum_{\T^k , A_k, \a_k \atop \b_1 ,\b_2 }  
 z_{12}^{\D^g (\T^k) - \D^g(\T^1) -\D^g (\T^2)} 
F_{A_1 A_2 \a_k}^{\b_1 \b_2 A_k} (\T^1 \T^2 \T^k)  
\left\{ 
 \left[ 1 + \frac{1}{2} \sum_{r=1}^{\infty} {z_{12}^r \over r !} \pa_2^r 
\right]   \gp (\T^k,z_2 )_{A_k}{}^{\a_k}  \d_{\b_1}^{\a_1}\d_{\b_2}^{\a_2}    
\right. 
$$
$$
+ \left.  \sum_{r=0}^{\infty} {z_{12}^{r+1} \over (r+1) !} \pa_2^r  
\{2L^{ab}_g :   \gp (\T^k,z_2)_{A_k}{}^{\a_k}  
\J_b (z_2) : \} \frac{1}{2} 
[(\T^1_a)_{\b_1}{}^{\a_1} \d_{\b_2}^{\a_2}    
-(\T^2_a) _{\b_2}{}^{\a_2} \d_{\b_1}^{\a_1}    ] \right\}  
$$
$$
+ \oko   
$$
which groups the Virasoro primary fields with their Virasoro 
descendants. For simplicity, we have omitted the $\oko$ terms which are
proportional to $:\gp \J \J :$ and derivatives thereof.  

\vs .4cm
\ni \un{Antichiral sector}
\vs .3cm

Following the same development for the product of two antichiral vertex
operators, we find:
\bs
\be
 \gm (\T^1,\bz_1) \gm (\T^2,\bz_2)   
= {\bz_{12}}^{2 L^{ab}_g \T_a^1 \T_b^2 }   
 : \gm (\T^1,\bz_1) \gm (\T^2,\bz_2) : 
 + \okt  
\ee
\be
\eqalign{
: \gm (\T^1,\bz_1)   \gm (\T^2,\bz_2) : &  \equiv  
\bG (\T^1 )   \bG (\T^2 ) \cr  
 -  i \T^1_a  \bG & (\T^1 )    \bG (\T^2 )  \bX^a(\bz_1) 
- i \T^2_a \bG (\T^1 )  \bG (\T^2 ) \bX^a(\bz_2) \cr  
 +   \T^1_a & \T^1_b\bG (\T^1 )  \bG (\T^2 )     \bN^{ab}(\bz_1) 
+ \T^2_a \T^2_b \bG (\T^1 )  \bG (\T^2 )   \bN^{ab}(\bz_2) \;\;\;\;\;\;\; \cr  
& + \T_a^1 \T_b^2 \bG (\T^1 )  \bG (\T^2 ) \bN_2^{ab} (\bz_1,\bz_2)  
+ \okt  \cr }  
\ee
\be
\eqalign{
 =  \bz_1^{-2\D^g(\T^1)} \bz_2^{- 2 \D^g(\T^2) }
 & (\bz_1 \bz_2)^{ - 2L^{ab}_g \T_a^1 \T_b^2 } \cr  
\ti & [  \one  -  i \T^1_a  \bX^a(\bz_1) - i \T^2_a \bX^a(\bz_2) \cr  
 & + \T^1_a \T^1_b    \bN^{ab}(\bz_1)  + \T^2_a \T^2_b   \bN^{ab}(\bz_2) \cr  
 & + \T_a^1 \T_b^2 \bN_2^{ab} (\bz_1,\bz_2)]  
\bG (\T^1 )  \bG (\T^2 ) + \okt  \cr}  
\ee 
\be
\bN_2^{ab} (\bz_1,\bz_2)  \equiv  N_2^{ab} (z_1,z_2) 
{\vert_{z_1 \ra \bz_1,z_2 \ra \bz_2, \J \ra \bJ}} 
\ee
\be
\bG(\T^1)_{\a_1}{}^{A_1} 
\bG(\T^2)_{\a_2}{}^{A_2}
=\sum_{ \T^k, A_k, \a_k}  
\bar{F}^{A_1 A_2 \a_k}_{\a_1 \a_2 A_k} (\T^1 \T^2 \T^k)  
	\bG(\T^k)_{\a_k}{}^ {A_k}  + \okt  
\label{agr} \ee 
\be
\bar{F}^{A_1 A_2 \a_k}_{\a_1 \a_2 A_k} (\T^1 \T^2 \T^k) 
= \bar{Q}^{A_1 A_2}{}_{A_k}(\T^1 \T^2 \T^k) 
\bC_{\a_1 \a_2 }{}^{\a_k}  (\T^1 \T^2 \T^k) + \okt 
\ee
\be
[ (\T_a^1)_{\a_1}{}^{\b_1} \d_{\a_2}{}^{\b_2} 
+\d_{\a_1}{}^{\b_1} (\T_a^2)_{\a_2}{}^{\b_2}] 
\bC_{\b_1 \b_2} {}^{\a_k}(\T^1 \T^2 \T^k) 
= 
\bC_{\a_1 \a_2} {}^{\b_k}(\T^1 \T^2 \T^k) 
(\T_a^k)_{\b_k}{}^{\a_k}  
\ee
\be
C^{\a \bar{\b} }{}_{.}  (\T \bar{\T} \T_{(1)})    
= {1 \over \sqrt{{\rm dim}\,\T}}\et^{\a \bar{\b}}  (\T)
\label{cmt} \ee
\be
Q_{A \bar{B}}{}^{.}  (\T \bar{\T} \T_{(1)})  
= {1 \over \sqrt{{\rm Tr}\,\L (\T) } } \L_{A \bar{B}}  (\T)
\label{qmt} \ee
\be
\bC_{\a_1 \a_2}{}^{\a_k} (\T^1 \T^2 \T^k) = \et_{\a_1 \b_1} (\T^1) 
\et_{\a_2 \b_2}(\T^2)  
C^{\b_1 \b_2 }{}_{\b_k} (\T^1 \T^2 \T^k)^* \et^{\b_k \a_k}  (\T^k)   
\label{dcg} \ee
\be
\bQ^{A_1 A_2}{}_{A_k} (\T^1 \T^2 \T^k) = \L^{A_1 B_1}(\T^1)
  \L^{A_2 B_2}(\T^2)  
Q_{B_1 B_2 }{}^{B_k} (\T^1 \T^2 \T^k)^* \L_{B_k A_k}  (\T^k)   
\ee
\be
\sum_{A_1 A_2} 
\bar{Q}^{A_1 A_2}{}_{A_k} (\T^1 \T^2 \T^k)   
Q_{A_1 A_2}{}^{A_l} (\T^1 \T^2 \T^l) 
= \d_{\T^k,\T^l} \d_{A_k}^{A_l} + \okt \pe  
\label{qsr} 
\ee
\es
Here $\T_{(1)}$ is the trivial representation, $\et_{\a \b}(\T)$ is the
carrier space metric of irrep $\T$ and $\L_{AB}(\T)$ is the corresponding
invariant form on the quantum group. $\bC$ and $\bQ$ are the duals of the
classical and quantum Clebsch-Gordan coefficients $C$ and $Q$.

After some algebra, we then obtain the OPE of two antichiral vertex operators, 
\be
\label{aop} 
\eqalign{ 
\gm & (\T^1,\bz_1)_{\a_1}{}^{A_1}   
\gm (\T^2,\bz_2)_{\a_2}{}^{A_2} 
\cr  
& = \sum_{\T^k , A_k, \a_k \atop \b_1}
 \bz_{12}^{\D^g (\T^k) - \D^g(\T^1) -\D^g (\T^2)} 
\left\{ 
   \gm (\T^k,\bz_2 )_{\a_k}{}^{A_k}  \d_{\a_1}^{\b_1}
 \phantom{  { \sum_{r}^{\infty} z_{2}^{r} \over s !}}  
\right. 
\cr
& - \sum_{r=0}^{\infty} {\bz_{12}^{r+1} \over (r+1) !}   
2L^{ab}_g :   \gm (\T^k,\bz_2)_{\a_k}{}^{A_k}    \bpa_2^r \bJ_b (\bz_2) : 
(\T^1_a)_{\a_1}{}^{\b_1}   \cr  
 & \left. +  \sum_{r,s=0}^{\infty} { \bz_{12}^{r+s+2} \over (r+s+2) !} 
       4 L^{ab}_g L^{cd}_g 
: \gm (\T^k,\bz_2)_{\a_k}{}^{A_k}  
\bpa_2^r [ \bpa_2^s \bJ_b(z_2) \bJ_c(z_2)] :   
(\T^1_a \T^1_d)_{\a_1}{}^{\b_1} \right\}  \cr    
&\;\;\;\;\;\;\;\;\;\;\;\;\;
 \ti \bF_{\b_1 \a_2 A_k}^{A_1 A_2 \a_k} (\T^1 \T^2 \T^k) 
+ \okt \cr}
\ee
and the antichiral analogue of (\ref{sop}) 
\be  
\gm (\T^1,\bz_1)_{\a_1}{}^{A_1}   
\gm (\T^2,\bz_2)_{\a_2}{}^{A_2} 
\phantom{ \C_{A_1 A_2 \a_k}^{\b_1 \b_2 A_k} (\T^1 \T^2 \T^k)  
 z_{12}^{\D^g (\T^k) - \D^g(\T^1) -\D^g (\T^2)}  
 z_{12}^{\D^g (\T^k) - \D^g(\T^1) -\D^g (\T^2)} } 
\ee
$$ 
= \sum_{ \T^k, A_k , \a_k \atop \b_1 , \b_2}  
 \bz_{12}^{\D^g (\T^k) - \D^g(\T^1) -\D^g (\T^2)} 
\left\{ 
 \left[ 1 + \frac{1}{2} \sum_{r=1}^{\infty} {\bz_{12}^r \over r !} \bpa_2^r 
\right]   \gm (\T^k,\bz_2 )_{\a_k}{}^{A_k}  \d_{\a_1}^{\b_1}\d_{\a_2}^{\b_2}   
\right. 
$$
$$
- \left.  \sum_{r=0}^{\infty} {\bz_{12}^{r+1} \over (r+1) !} \bpa_2^r  
\{2L^{ab}_g :   \gm (\T^k,\bz_2)_{\a_k}{}^{A_k}  \bJ_b (\bz_2) : \} \frac{1}{2} 
[(\T^1_a)_{\a_1}{}^{\b_1} \d_{\a_2}^{\b_2}    
-(\T^2_a) _{\a_2}{}^{\b_2} \d_{\a_1}^{\b_1}    ]\right\}  
$$
$$
\ti \bF_{\b_1 \b_2 A_k}^{A_1 A_2 \a_k} (\T^1 \T^2 \T^k) + \oko  \pe  
$$
is also obtained with the high-level ODE (\ref{oda}).  

\vs .4cm
\ni \un{PDE's for operator products}
\vs .3cm

Using the algebra (\ref{egp},b) and the
 ODE's (\ref{cod},b)  for $g_{\pm}$, a set of exact
 PDE's for the product of
two vertex operators are easily derived \cit{26}:
\bs
\be
P_+ (z_1,z_2)_{A_1 A_2}{}^{\a_1 \a_2} \equiv 
\gp (\T^1,z_1)_{A_1}{}^{\a_1} 
\gp (\T^2,z_2)_{A_2}{}^{\a_2} 
\;\;\;\;\;\;\;\;\;\;\;\;\;\;\;\;\;\;\;\;\;\;\; \;\;\;\;\;\;\;\;\;\;\;\;\;\;\;\;
\ee
\be
\pa_1 P_+ (z_1,z_2)
= 2L^{ab}_g \{: P_+ (z_1,z_2)\J_a(z_1) : \T_b^1 + 
{ P_+ (z_1,z_2) \T_a^1 \T_b^2 \over z_{12} } \} 
\ee
\be
\pa_2 P_+ (z_1,z_2)
= 2L^{ab}_g \{: P_+ (z_1,z_2)\J_a(z_2) : \T_b^2  - 
{ P_+ (z_1,z_2) \T_a^1 \T_b^2 \over z_{12} } \} 
\ee
\be
P_- (\bz_1,\bz_2)_{\a_1 \a_2}{}^{A_1 A_2}
\equiv 
\gm (\T^1,\bz_1)_{\a_1}{}^{A_1}  
\gm (\T^2,\bz_2)_{\a_2}{}^{A_2}
\;\;\;\;\;\;\;\;\;\;\;\;\;\;\;\;\;\;\;\;\;\;\; \;\;\;\;\;\;\;\;\;\;\;\;\;\;\;\;
\;\;
\ee
\be
\bpa_1 P_- (\bz_1,\bz_2)
=  - 2L^{ab}_g \{ \T_b^1 : P_- (\bz_1,\bz_2)\bJ_a(\bz_1)  : -  
{\T_a^1 \T_b^2  P_- (\bz_1,\bz_2) \over \bz_{12} } \} 
\ee
\be
\;\;\;\;\;\;\;\;\;\;\;\;\;\; \bpa_2 P_- (\bz_1,\bz_2)
=  - 2L^{ab}_g \{\T_b^2  : P_- (\bz_1,\bz_2)\bJ_a( \bz_2) :  +  
{\T_a^1 \T_b^2 P_- (\bz_1,\bz_2) \over \bz_{12} } \} \pe  
\ee
\es
The normal ordering here is the same as in (\ref{nop}) with $g \ra P$, and
we have checked explicitly that the OPE's (\ref{cop}) and (\ref{aop}) satisfy
these PDE's thru the appropriate order.

\vs .4cm
\ni \un{Braid relations}
\vs .3cm

We close this section with a short discussion of braid relations which provides
a check  of our formulation against standard relations in the literature. 

The braid matrix (or universal $R$ matrix) $\B_g$, which acts on the quantum
group indices of the chiral vertex operator is defined by the braid relation 
\bs
\be
\gp (\T^1,z_1)_{A_1}{}^{\a_1} \gp (\T^2,z_2)_{A_2}{}^{\a_2}   
= \B_g (\T^1\T^2)_{A_1 A_2}{}^{B_2 B_1}  
 \gp (\T^2,z_2)_{B_2}{}^{\a_2}  \gp (\T^1,z_1)_{B_1}{}^{\a_1}   
\ee
\be 
\B_g (\T^2\T^1) 
= \B_g ^{-1} (\T^1\T^2) \pe  
\label{ibm} \ee
\label{bdf} \es 
Using (\ref{bdf}) and (\ref{acf}) in the OPE (\ref{cop}), we find
that the braid matrix also describes
the $1 \lra 2$ exchange of the fusion coefficient $F$ of the zero modes
in (\ref{ggr}), 
\be
\eqalign{
& F_{A_1 A_2 \a_3}^{\a_1 \a_2 A_3} (\T^1 \T^2 \T^3) \cr  
 &= \B_g(\T^1 \T^2)_{A_1 A_2}{}^{B_2 B_1} 
F_{B_2 B_1 \a_3}^{\a_2 \a_1 A_3} (\T^2 \T^1 \T^3) 
{\rm e}^{-i \pi [\D^g(\T^3) -  \D^g(\T^1) -  \D^g(\T^2)]   
{\rm sign (arg} (z_1/z_2) )} 
+ \okt \; .  \cr}
\ee
Using next the factorized form of $F$ in (\ref{fff}) and the exchange relation 
of the Clebsch-Gordan coefficient $C$, 
\bs
\be
C^{\a_2 \a_1 }{}_{\a_3}  (\T^2 \T^1 \T^3) 
= \n (\T^1 \T^2 \T^3) C^{\a_1 \a_2 }{}_{\a_3} (\T^1 \T^2 \T^3) 
 \ee
\be
\n (\T^1 \T^2 \T^3) =\n (\T^2 \T^1 \T^3) = \pm 1
\ee
\es
we see that the braid matrix also describes the $1 \lra 2 $ exchange of
the quantum Clebsch-Gordan coefficients $Q$ 
\be
\label{qgr}
\eqalign{
 Q_{A_1 A_2}{}^{A_3}  (\T^1 \T^2 \T^3) 
 = & \B_g (\T^1 \T^2) _{A_1 A_2}{}^{B_2 B_1} 
Q_{B_2 B_1}{}^{A_3} (\T^2 \T^1 \T^3) \cr  
\ti  &\n (\T^1 \T^2 \T^3)  
{\rm e}^{-i \pi [\D^g(\T^3) -  \D^g(\T^1) -  \D^g(\T^2)]   
{\rm sign (arg} (z_1/z_2) )} 
+ \okt \pe \cr}
 \ee
To discuss this relation further, we introduce the invariant level
$x= 2 k /\ps_g^2$ of the affine algebra, where $\ps_g$ is the
highest root of $g$.   
When   ${\rm arg}(z_2/z_1)  >0$ and the quantum group parameter $q$ is
identified  as usual as  
\be
q = {\rm e}^{ \frac{2 \pi i}{ x  } } 
+ \okt  
\ee
we find that the relation (\ref{qgr}) agrees with the high-level form of the 
corresponding quantum $SU(2)$ relation 
\be
K_{j}^{j_1 j_2} R^{j_2 j_1} 
= (-1)^{j_1 +j_2 -j} q^{(c_j -c_{j_1} -c_{j_2})/2} K_j^{j_2 j_1} 
\sp c_j = j(j+1) \sp q= \exp \left( {2 \pi i \over x+2} \right )   
\pe
\ee
which appears as eq.(13) in Ref.\cit{30}. 
To see this agreement, use the $SU(2)$ identifications
\bs
\be
\B_g (\T^1 \T^2) \ra  R^{j_1 j_2} 
\sp \D_g (\T) \ra {j(j+1) \over x+2} 
\ee
\be
Q (T^1 T^2 T^3 ) \ra  K_{j}^{j_1 j_2} 
\sp 
\nu (\T^1 \T^2 \T^3) \ra (-1)^{j_1 + j_2-  j} 
\ee
\es
and the inverse relation (\ref{ibm}) to move $\B_g$ to the left of
(\ref{qgr}).

\section{Chiral Correlators}

Using the semiclassical vertex operators (\ref{cvo}) and the algebra 
(\ref{gza}),
one straightforwardly computes the semiclassical chiral correlators (\ref{cco}),
\be
A_g^+ (\T,z)
  = \langle 0 |  \pG (\T^1) \cdots  \pG (\T^n)  | 0 \rangle    
\left[ 1 + 2 L^{ab}_g \sum_{i <j}^n\T_a^i \T_b^j  \ln z_{ij} 
\right] + \okt 
\label{dcc} \ee
up to the constant zero-mode averages
  $ \langle  \pG \cdots  \pG    \rangle   $. This simple result is obtained
because the normal-ordered terms
in (\ref{cvo}) once again fail to contribute to the correlators, and the result
(\ref{dcc}) indeed satisfies the chiral KZ equations (\ref{hlk}). 
Using the algebra (\ref{zmc}), one finds that the zero-mode averages, and hence
the chiral correlators (\ref{dcc}), satisfy the $g$-global Ward identities
\bs
\be 
   \langle 0 |  \pG (\T^1) \cdots  \pG (\T^n)  | 0 \rangle    
\sum_{i=1}^n \T_a^i = 0 + \okt  
\ee
\be
A_g^+(\T,z) 
\sum_{i=1}^n \T_a^i  = 0 + \okt     
\ee
\es
so that these quantities are $g$-invariant thru this order of the
semiclassical expansion. 
As discussed in Ref.\cit{25}, the chiral correlators, and hence the zero-mode
averages, are similarly invariant under the quantum group. 

To be more explicit about the zero-mode averages, we may use the  
fusion relation (\ref{ggr}) to obtain the forms,
\bs
\be
\langle 0 |\pG (\T)_{A}{}^{\a} |0\rangle =\d(\T,\T_{(1)} )   
\;\;\;\;\;\;\;\;\;\;\;\;\;\;\;\;
\;\;\;\;\;\;\;\;\;\;\;\;\;\;\;\;
\;\;\;\;\;\;\;\;\;\;\;\;\;\;\;\;
\;\;\;\;\;\;\;\;\;\;\;\;\;\;\;\;
\;\;\;\;
\;\;
\ee
\be
\langle 0 |\pG (\T^1)_{A_1}{}^{\a_1} \pG (\T^2)_{A_2}{}^{\a_2} 
|0\rangle =
\d(\T^2,\bar{\T}^1 ) F^{\a_1 \a_2 . }_{A_1 A_2 . }  (\T^1 \bar{\T}^1 \T_{(1)})
+ \okt  
\;\;\; 
\ee
\be
\eqalign{
\langle 0 & |\pG (\T^1)_{A_1}{}^{\a_1} \pG (\T^2)_{A_2}{}^{\a_2}
\pG (\T^2)_{A_3}{}^{\a_3}  
|0\rangle  \cr
& = F^{\a_1 \a_2 \bA_3 }_{A_1 A_2 \bar{\a}_3 }  (\T^1 \T^2 \bar{\T}^3 ) 
F^{\bar{\a}_3 \a_3 . }_{\bA_3 A_3 .  }  (\bar{\T}^3 \T^3 \T_{(1)} ) 
+ \okt 
\;\;\;\;\;\;\;\;\;\;\;\;\;\;\;\;
\;\;\;\;\;\;\;\;\;\;\;\;\;\;\;\;
 \cr }
\ee
\be
\eqalign{
\;\;  \langle 0  & |  
\pG (\T^1)_{A_1}{}^{\a_1}  
\pG(\T^2)_{A_2}{}^{\a_2}   \pG(\T^3)_{A_3}{}^{\a_3}   
\pG (\T^4)_{A_4}{}^{\a_4}  | 0 \rangle   \cr  
 & = \sum_{\T^k  {A_k , \bA_k \atop  \a_k,  \bar{\a}_k }}  
F_{A_1 A_2 \a_k}^{\a_1 \a_2 A_k} (\T^1 \T^2 \T^k)  
F_{A_3 A_4 \bar{\a}_k}^{\a_3 \a_4 \bA_k} (\T^3 \T^4 \bar{\T}^k)  
F_{A_k \bA_k .}^{\a_k \bar{\a}_k .} (\T^k \bar{\T}^k \T_{(1)}) + \okt \;\; \cr} 
\ee 
\es
where $\d$ is Kronecker delta and $\T_{(1)}$ is the trivial representation.
Using (\ref{fff}), the zero-mode averages can also
 be expressed in terms of (level-independent) classical and (level-dependent) 
quantum group invariants $\bv$ and $d$ 
\bs
\be
\langle 0 |  
\pG (\T^1)
\cdots \pG (\T^n)  | 0 \rangle_A{}^\a    
= \sum_{m } d_{A}^m  \bv_m^{\a}  + \okt  
\ee
\be
v_\a^m \bv_m^\b = (I_g^n)_\a{}^\b   
\sp
\bv_m^\a   v_\a^l  = \d_m^l   
\label{onr} \ee
\be
d_A^m \bd_m^B = (D_g^n)_A{}^B 
\sp
\bd_m^A d_A^l = \d_m^l 
\ee
\label{npa} \es
where we have also introduced the dual invariants $v$ and $\bd$ to write
the completeness and orthonormality relations (\ref{onr},c) of the 
invariants. The
quantity $D_g^n$ is the quantum analogue of the classical invariant
projector $I_g^n$.
  
In further detail, we find for the four-point average
\bs
\be
\langle 0 |  
\pG (\T^1)_{A_1}{}^{\a_1}  
\pG(\T^2)_{A_2}{}^{\a_2}   \pG(\T^3)_{A_3}{}^{\a_3}   
\pG (\T^4)_{A_4}{}^{\a_4}  | 0 \rangle    
= \sum_m d({\rm s},g)_{A}^m  \bv ({\rm s},g)_m^{\a} + \okt  
\ee
\be
 \bv (s,g)_m^\a  \equiv {1 \over \sqrt{{\rm dim}\,\T^m } }   
\sum_{\a_m , \bar{\a}_m  } 
C^{\a_1 \a_2}{}_{\a_m} (\T^1 \T^2 \T^m)  
C^{\a_3 \a_4}{}_{\bar{\a}_m} (\T^3 \T^4 \bar{\T}^m)  
\et^{\a_m \bar{\a}_m}  (\T^m)
\label{vsg} \ee
\be
 d (s,g)_{A}^m  \equiv {1 \over \sqrt{  {\rm Tr}\,\L (\T^m) }}  
\sum_{A_m , \bA_m  } 
Q_{A_1 A_2}{}^{A_m} (\T^1 \T^2 \T^m)  
Q_{A_3 A_4}{}^{\bA_m } (\T^3 \T^4 \bar{\T}^m)  
\L_{A_m \bA_m }(\T^m) 
\ee
\label{bex} \es
where $C$ and $Q$ are the classical and quantum Clebsch-Gordan coefficients,
and $\et_{\a \b}(\T)$ and  $\L_{AB}(\T)$ are  
defined in (\ref{cmt}) and (\ref{qmt}).
The form (\ref{vsg}) of $\bv(s,g)_m$ was first given in Ref.\cit{42}, 
where these 
quantities are called the s-channel invariants of the four-point correlator.
Similarly, the quantities $d(s,g)^m$ may be interpreted as the s-channel 
quantum invariants of the 4-point correlator. 

The classical and quantum invariants $\bv$ and $d$
also appear in the conformal-block
expansion of the chiral correlators. Using the KZ gauge
\bs 
\be
A_g^+ (z_1,z_2,z_3,z_4)_A{}^\a 
= { Y_g^+ (y)_A{}^\a \over  \prod_{i <j}^4 z_{ij}^{\gamma^g_{ij}} } 
\sp y  ={z_{12} z_{34} \over z_{14} z_{32} }
\ee
\be
  \gamma^g_{12} =  
  \gamma^g_{13} = 0  
\sp
  \gamma^g_{14} = 2 \D^g (\T^1)   
\sp
  \gamma^g_{23} =  
\D^g (\T^1)+   \D^g (\T^2)+   \D^g (\T^3) -  \D^g (\T^4)   
\ee
\be
  \gamma^g_{24} =  
- \D^g (\T^1)+   \D^g (\T^2)-   \D^g (\T^3) +  \D^g (\T^4)   
\;,\; 
  \gamma^g_{34} =  
- \D^g (\T^1)-  \D^g (\T^2)+   \D^g (\T^3)  + \D^g (\T^4)   
\ee
\es
and the $g$-global Ward identity in (\ref{ggi}), we find from (\ref{dcc})
and (\ref{bex}) that
\bs
\be
Y_g^+ (y)_A{}^\a  = 
\sum_{m,l} d({\rm s},g)_{A}^m \F_g^{({\rm s})} (y)_m{}^l  
 \bv ({\rm s},g)_l^{\a} + \okt  
\label{bld} \ee
\be
 \F_g^{({\rm s})} (y)_m{}^l  
=  \bv ({\rm s},g)_m^{\a}   
[ \one + 2L^{ab}_g ( \T_a^1 \T_b^2 \ln y + 
 \T_a^1 \T_b^3 \ln (1-y) ) ]_\a{}^\b   
 v ({\rm s},g)^l_{\b} + \okt   
\ee
\label{asb} \es
where $\F_g^{(s)}$ are the semiclassical s-channel affine-Sugawara blocks
obtained in Ref.\cit{42}. The reproduction of the affine-Sugawara blocks
in (\ref{asb}) is a central check on the new nonabelian chiral vertex 
operators (\ref{vop}). We emphasize however that,  
in the conventional block analysis [7,42]  of the solutions
of the KZ equations, the quantum invariants $d_A^m$ in (\ref{bld}) are 
discarded as irrelevant constants. 

The corresponding results for the antichiral correlators (\ref{aco}) are,  
\bs
\be
 A_g^-(\T,\bz) = 
\left[ 1 + 2 L^{ab}_g \sum_{i <j}^n\T_a^i \T_b^j  \ln \bz_{ij} \
\right] 
\langle 0 |  \bG (\T^1) \cdots  \bG (\T^n)  | 0 \rangle   + \okt  
\ee
\be
\langle 0 |  
\bG (\T^1)
\cdots \bG (\T^n)  | 0 \rangle_\a{}^A     
= \sum_{m } v_{\a}^m   \bd_m^{A}   +\okt  
\label{anp} \ee 
\be
A_g^- (\bz_1,\bz_2,\bz_3,\bz_4)_\a{}^A  
= { Y_g^- (\by)_\a{}^A  \over  \prod_{i <j}^4 \bz_{ij}^{\gamma^g_{ij}} } 
\ee
\be
Y_g^- (\by)_\a{}^A = \sum_{m,l} v ({\rm s},g)_\a^m \F_g^{({\rm s})} (\by)_m{}^l 
\bd ({\rm s},g)_l^A + \okt  
\ee
\be 
 \F_g^{({\rm s})}  (\by)_m{}^l   = (\F_g^{({\rm s})} (y)_l{}^m)^* 
\ee
\es
where $v$ and $\bar{d}$ are the dual invariants to $\bv$ and $d$.

\vs .4cm
\ni \un{Braiding and crossing}
\vs .3cm
 
The braid matrix $\B_g$ in \cit{25} 
\be 
\gp (\T^2,z_2)_{A_2}{}^{\a_2} \gp (\T^3,z_3)_{A_3}{}^{\a_3}  
= \B_g (\T^2\T^3)_{A_2 A_3}{}^{B_3 B_2}  
 \gp (\T^3,z_3)_{B_3}{}^{\a_3}  \gp (\T^2,z_2)_{B_2}{}^{\a_2}   
\label{brr} \ee
describes the exchange of two chiral vertex operators, while the s-u 
crossing matrix $X_g({\rm su})$ in Ref.\cit{42}
\bs
\be
\F_g^{({\rm s})}(y)_m{}^l = X_g({\rm su})_m{}^p \F_g^{({\rm u})}(y)_p{}^q   
X_g^{-1} ({\rm su})_q{}^l +\ok  
\ee
\be
\bv ({\rm s},g)_m =  X_g ({\rm su})_m{}^l \bv ({\rm u},g)^l +\ok  
\ee 
\be
X_g ({\rm su})_m{}^l = \bv ({\rm s},g)_m v ({\rm u},g)^l +\ok  
\ee
\label{crr} \es
relates the s and u channel conformal blocks $\F_g^{({\rm s})}$ and
$\F_g^{({\rm u})}$ of the affine-Sugawara construction. 
It is clear that $\B_g$ and $X_g$ 
represent the same physical operation, namely the exchange of external states
$2 \lra 3$. The braid matrix acts however in the quantum
space with $A_i,B_i =1 \ldots
{\rm dim}\,\T^i$, while the crossing matrix acts on the generally smaller space
of conformal blocks $\F_g$ with $m,l = 1 \ldots {\rm dim}({\rm invariants})$.
It follows that there must be a map, or intertwining relation, between
the braid matrix $(\B_g)_A{}^B$ and the crossing matrix $(X_g)_m{}^l$ and 
we shall 
find that the intertwiner is the set of 
quantum invariants $d_A^m$.

To see this, we first
translate the braid relation (\ref{brr}) into the braid relation
of the chiral correlator $A_g^+$
\be
A_g^+(z_1,z_2,z_3,z_4)_{A_1 A_2 A_3 A_4}^{\a_1 \a_2 \a_3 \a_4}   
= \B_g (\T^2\T^3)_{A_2 A_3}{}^{B_3 B_2}  
A_g^+(z_1,z_3,z_2,z_4)_{A_1 B_3 B_2 A_4}^{\a_1 \a_3 \a_2 \a_4} \pe  
\ee
In terms of the invariant four-point correlator $Y_g^+$, this relation reads 
\be
\eqalign{
 Y_g^+ & (y)_{A_1 A_2 A_3 A_4}^{\a_1 \a_2 \a_3 \a_4} \cr  
  = & \B_g (\T^2\T^3)_{A_2 A_3}{}^{B_3 B_2}  
Y_g^+(1-y)_{A_1 B_3 B_2 A_4}^{\a_1 \a_3 \a_2 \a_4} 
 {\rm e}^{ i \pi ( \D^g(\T^1) + \D^g(\T^2) + \D^g(\T^3) - \D^g(\T^4) )  
 {\rm sign (arg} (z_2/z_3) )    } \cr  
& \;\;\;\;\;\;\;\;\;\;\;\;\; + \okt  \pe \cr}  
\ee
Then using the block decomposition (\ref{bld}) along with the similar
u-channel decomposition 
\be
Y_g^+ (1-y)_{A_1 A_3 A_2 A_4}{}^{\a_1 \a_3 \a_2 \a_4}  = 
\sum_{m,l} d({\rm s},g)_{A_1 A_3 A_2 A_4}^m \F_g^{({\rm u})} (y)_m{}^l  
 \bv ({\rm u},g)_l^{\a_1 \a_2 \a_3 \a_4} + \okt  
\ee
 we obtain the intertwining relation
\be
 \eqalign{
 \B_g & (\T^2\T^3)_{A_2 A_3}{}^{B_3 B_2} d({\rm s},g)_{A_1 B_3 B_2 A_4}^m  \cr  
  = &  
\sum_l d({\rm s},g)_{A_1 A_3 A_2 A_4}^l X_g (su)_l{}^m 
{\rm e}^{- i \pi [ \D^g(\T^1) + \D^g(\T^2) + \D^g(\T^3) - \D^g(\T^4) ]   
{\rm sign (arg} (z_2/z_3) )    } + \okt  
\pe  \cr } 
\ee  
 Here, we have also used the crossing relations (\ref{crr}) and completeness
of the $g$-invariants $\bv({\rm s},g)$. 
For the special case of certain irreps of $SU(n)$, this relation was
obtained exactly in Ref.\cit{40}.

\section{Chiral-Antichiral OPE's} 

We turn next to the question
of the chiral-antichiral OPE's, which are known from the
action formulation [34,35,37,40,41]  to
depend on the treatment of the constant quantum space ambiguity in
the factorization 
$ g(\T,\bz,z)_\a{}^\b = \gm (\T,\bz)_\a{}^A \gp (\T,z)_A{}^\b$.  
In what follows
we briefly discuss this ambiguity, as it is reflected in our operator
formulation.

We begin by assuming the exact operator solution
\be
[\J_a(m),  \gm (\T,\bz)] = 
 [\bJ_a(m),  \gp (\T,z)] = 0 
\label{ops} \ee
of the conditions (\ref{agb}). The choice (\ref{ops}) is the natural solution
of (\ref{agb}) because we now know  that 
$\gp$ and $\gm$ 
are unitary (and hence invertible) in the extreme semiclassical limit.
Moreover, (\ref{ops}) immediately implies the natural relations 
\be
[ \bT_g (\bz) , \gp (\T,z) ] = [T_g (z)  , \gm (\T,\bz) ] = 0 
\ee 
which say that the chiral and antichiral vertex operators are inert under
the antichiral and chiral affine-Sugawara constructions respectively. 

Using (\ref{ops}),  one can also derive the regular PDE's
\bs
\be
P_{+-} (z_1,\bz_2)_{A_1 \a_2}{}^{\a_1 A_2} \equiv 
\gp (\T^1,z_1)_{A_1}{}^{\a_1} 
\gm (\T^2,\bz_2)_{\a_2}{}^{A_2} 
\;\;\;\;\;\;\;\;\;\;
\ee
\be
\pa_1 P_{+-} (z_1,\bz_2)
= 2L^{ab}_g : P_{+-}  (z_1,\bz_2)E_a(z_1)  : \T_b^1 
\ee
\be
\;\;\;\, \bpa_2 P_{+-} (z_1,\bz_2)
= - 2L^{ab}_g : P_{+ -}(z_1,\bz_2) \bJ_a(\bz_2)   : \T_b^2  
\ee
\be
P_{-+} (\bz_2,z_1)_{\a_2 A_1}{}^{A_2 \a_1} \equiv 
\gm (\T^2,\bz_2)_{\a_2}{}^{A_2} 
\gp (\T^1,z_1)_{A_1}{}^{\a_1} 
\;\;\;\;\;\;\;\;\;\;
\ee
\be
\pa_1 P_{-+} (\bz_2,z_1)
= 2L^{ab}_g : P_{-+}  (\bz_2,z_1)E_a(z_1)  : \T_b^1 
\ee
\be
\;\;\;\, \bpa_2 P_{-+} (\bz_2,z_1)
= - 2L^{ab}_g : P_{ -+}(\bz_2,z_1) \bJ_a(\bz_2)   : \T_b^2  
\ee
\es
for the chiral-antichiral products, and these differential equations
tell us immediately  that the chiral-antichiral operator products
and OPE's are regular,
\bs
\be
\gp (\T^1,z_1) \gm (\T^2,\bz_2) = \mbox{regular in } (z_1 -\bz_2) 
\ee
\be 
\;\;\;\;\gm (\T^2,\bz_2) \gp (\T^1,z_1)= \mbox{regular in } (\bz_2 -z_1) 
\pe  
\ee
\label{cao} \es
Because of possibly non-trivial braiding however, the regularity of
these OPE's does not necessarily imply that the chiral and antichiral
vertex operators commute.

In our semiclassical expansion, one can be more explicit about
the chiral-antichiral operator products. We first need the chiral-antichiral 
commutators of the currents with the zero modes 
\bs
\be
[\J_a(0),  \Gm (\T)] = 
 [\bJ_a(0),  \Gp (\T] = 0 + \okt 
\ee
\be
[\J_a(m \neq 0),  \Gm (\T)] = 
 [\bJ_a(m \neq 0),  \Gp (\T] = 0 + \oko  
\ee
\label{mza} 
\es 
which follow from (\ref{ops}) and the expressions (\ref{icv}), (\ref{iav}) 
for the zero modes in terms of the primary fields. 
The results (\ref{gza}), (\ref{aza}) and (\ref{mza}) collect the complete 
semiclassical
algebra of the currents with the chiral and antichiral zero modes. 

The commutators (\ref{mza}) are the natural solutions
(because $\Gp(\T)$ and $\Gm (\T)$ are also unitary in the extreme semiclassical 
limit) of the general chiral-antichiral conditions 
\bs
\be
[\J_a(0),  \Gm (\T)] \Gp (\T) =0 + \okt \sp  
[\J_a(m \neq 0),  \Gm (\T)] \Gp (\T) =0 + \oko 
\label{ago} \ee
\be
\Gm (\T) [\bJ_a(0),  \Gp (\T)] = 0 +\okt 
\sp \Gm (\T) [\bJ_a(m \neq 0 ),  \Gp (\T)] = 0 +\oko  
\ee
\label{agz} 
\es
which are the zero-mode analogues of the general conditions (\ref{agb}).
The relations (\ref{ago},b) follow directly from (\ref{agb}), without 
the assumption (\ref{ops}), using
only the current algebra and the inversions (\ref{icv}) and (\ref{iav}).

Continuing with (\ref{mza}), and following the steps of Section 5,
one obtains the chiral-antichiral operator products as follows,
\bs
\be
  \gp (\T^1,z_1)_{A_1}{}^{\a_1}  \gm (\T^2,\bz_2)_{\a_2}{}^{A_2} 
 = \Gp (\T^1)_{A_1}{}^{\b_1}   \Gm (\T^2)_{\b_2}{}^{A_2}  
 M (z_1,\bz_2)_{\b_1 \a_2} {}^{\a_1 \b_2}  + \okt  
\ee
\be
 \gm (\T^2,\bz_2)_{\a_2}{}^{A_2} 
  \gp (\T^1,z_1)_{A_1}{}^{\a_1} 
 = \Gm (\T^2)_{\b_2}{}^{A_2}  \Gp (\T^1)_{A_1}{}^{\b_1}   
 M (z_1,\bz_2)_{\b_1 \a_2} {}^{\a_1 \b_2}  + \okt  
\ee
\be
  [ \gp (\T^1,z_1)  , \gm (\T^2,\bz_2)] 
 = [\Gp (\T^1),  \Gm (\T^2)]  M (z_1,\bz_2) + \okt  
\ee
\label{cwo} 
\es
where the explicit form of $M(z_1,\bz_2)$ is 
\be
\label{mfo}
\eqalign{
M(z_1,\bz_2) = \one & + i X^a(z_1) \T_a^1 - i \bX^a(\bz_2) \T_a^2
+ X^a (z_1) \bX^b(\bz_2) \T_a^1 \T_b^2 \cr
& + N^{ab} (z_1) \T_b^1 \T_a^1   +  
\bN^{ab} (\bz_2) \T_a^2 \T_b^2  + \okt \pe  }
\ee
The form (\ref{mfo}) for $M(z_1,\bz_2)$ can easily be expanded to find the 
explicit
semiclassical forms of the regular OPE's (\ref{cao}).

The result  (\ref{cwo}) shows that the ambiguity of the 
chiral-antichiral OPE's, and hence the chiral-antichiral commutator, is entirely
in the constant zero-mode products $ \Gp (\T) \Gm (\T)$ and $\Gm(\T) \Gp(\T)$,  
which then carry the constant quantum space ambiguity [35,40,41]   
of the factorization. Among the various
possibilities however, the most esthetic solution is the one in which
the chiral and antichiral sectors commute
\bs
\be
\label{gac}
   [\Gp (\T^1) , \Gm (\T^2) ]  = 0 + \okt 
\;\;\;\;\;\;\;\;\;\;\;\;\;\;\;
\ee
\be   
   \Rightarrow [\gp (\T^1,z) , \gm (\T^2,\bz) ] = 0 + \okt 
\ee
\label{gch} \es
which is the ``gauge choice'' discussed by Caneschi and Lysiansky in 
Ref.\cit{41}.
In this case, we also obtain the algebra
\be
  [\Gp (\T^1) , \gm (\T^2,\bz) ] 
 =  [\gp (\T^1,z) , \Gm (\T^2) ]  
= 0 + \okt 
\ee
of the chiral zero modes with the antichiral vertex operators and vice versa. 
Unless otherwise stated, we limit the discussion below to the gauge choice 
(\ref{gch}).

\section{Semiclassical WZW Vertex Operators}

Our final task is to assemble the semiclassical chiral and antichiral sectors
into semiclassical WZW theory, beginning with the
semiclassical WZW vertex operators. Using (\ref{fas}),  
 (\ref{cvo}) and (\ref{avo}) we find  
\bs
\be 
g(\T,\bz,z)_\a{}^\b  = \gm (\T,\bz)_\a{}^A \gp (\T,z)_A{}^\b 
\phantom{\T_a \T_b \bq^{ab} (\bz) ]_\a{}^\r  G (\T)_\r{}^\s   
  + \okt\T_a \T_b \bq^{ab} } \;\;\;\;\;\;\;      
\ee
\be
= \bz^{-2\D^g (\T) }
 [ \one   -  i \T_a   \bX^a (\bz)   +
\T_a \T_b \bN^{ab} (\bz) ]_\a{}^\r  G (\T)_\r{}^\s   
 [ \one + i X^a(z)  \T_a + N^{ab}(z) \T_b \T_a ]_\s{}^\b
  + \okt  
\label{wvo} \ee 
\be
G (\T)_\a{}^\b  \equiv  
\sum_A \bG (\T)_\a{}^A \pG (\T)_A  {}^\b
\label{gdf} \ee
\be
G(\T) = \cO (k^0) 
\ee
\be
\pa G(\T) = \bpa G(\T) = 0 + \okt 
\ee 
\es
where $G (\T)$ in (\ref{gdf}) is the WZW zero mode.

Here are some important properties of the WZW zero mode.

\ni {\bf A.} Semiclassical unitarity. The
 extreme semiclassical unitarity of $G (\T) $,
\be
G^\dagger (\T) G (\T) = \one + \oko 
\label{scu} \ee
follows from the extreme semiclassical unitarity of $\Gp (\T)$ and $\Gm (\T)$.
This property is independent of the gauge choice (\ref{gch}).

\ni {\bf B.} Algebra with the currents. 
The  algebra of the currents with the WZW zero mode 
\bs 
\be
\label{eal}
[\J_a (0) , G (\T) ] = G (\T)   \T_a + \okt 
\;\;\;\;\;\;\;\;\;\;\;\;\;\;\;\;\;\;\;
 \ee
\be
[\J_a (m \neq 0 ) , G (\T) ] =  {i \over 2 k m } : G (\T)
\J_b(m):  f_{a}{}^{bc}\T_c         
+ \oko  
\ee
\be 
[ \bJ_a (0) , G (\T) ] = - \T_a G(\T)  + \okt 
\;\;\;\;\;\;\;\;\;\;\;\;\;\;\;
\ee
\be 
\;\;
[ \bJ_a (m \neq 0 ) , G (\T)] =  -{i  \over 2k m}  f_a{}^{bc} \T_c 
: G (\T) \bJ_b(m) :  
+ \oko  
\ee
\label{gal} \es
 follows easily from (\ref{gza}), (\ref{aza}) and the gauge choice 
(\ref{gac}).
Curiously, these relations can also be derived from (\ref{gza}), (\ref{aza}) and
the general conditions (\ref{agz}), so they are in fact also independent of the 
gauge choice (\ref{gch}). The algebra (\ref{gal}) is consistent with the 
semiclassical unitarity of $G(\T)$ in (\ref{scu}).   

\ni {\bf C.} Fusion and group multiplication. 
The fusion rule for two WZW zero modes,
\bs
\be
G(\T^1)_{\a_1}{}^{\b_1} 
G(\T^2)_{\a_2}{}^{\b_2} 
=\sum_{\T_k , \a_k, \b_k }  
D_{\a_1 \a_2 \b_k}^{\b_1 \b_2 \a_k} (\T^1 \T^2 \T^k)  
G(\T^k)_{\a_k}{}^{\b_k} +\okt  
\ee
\be
 D_{\a_1 \a_2 \b_k}^{\b_1 \b_2 \a_k} (\T^1 \T^2 \T^k)  
= \bC_{\a_1 \a_2}{}^{\a_k} (\T^1 \T^2 \T^k)     
C^{\b_1 \b_2 }{}_{\b_k} (\T^1 \T^2 \T^k)  
\ee
\label{gfr} \es
follows from (\ref{gac}), (\ref{ggr}), (\ref{fff}), (\ref{agr},f) and 
(\ref{qsr}), 
where $C$ and $\bC$ are the classical Clebsch-Gordan coefficients and their 
duals defined in (\ref{dcg}). To this order of the semiclassical expansion,
the fusion rule (\ref{gfr}) is the same as the multiplication law for 
classical group elements in the Clebsch basis.

\ni {\bf D.} Averages. 
Using (\ref{gac}), (\ref{npa}) and (\ref{anp}), we find that the zero-mode 
averages are Haar integrals
\be
\label{wav}
\eqalign{ 
\langle 0| G(\T^1) \cdots G(\T^n) | 0 \rangle_\a{}^\b   
& = \langle 0| \Gm(\T^1) \cdots \Gm(\T^n) | 0 \rangle_\a{}^A   
\langle 0| \Gp(\T^1) \cdots \Gp(\T^n) | 0 \rangle_A{}^\b   
\cr 
& = \sum_{m,l,A} v_\a^m \bd_m^A d_A^l \bv_l^\b +\okt    
 = \sum_{m} v_\a^m  \bv_m^\b  +\okt  \cr 
& = I_g^n + \okt  
  = \int {\rm d} \G \, \G (\T^1)  \cdots \G(\T^n) + \okt \cr}  
\ee
thru the indicated order. This result is also consistent with the
algebra (\ref{gal}).
 
Using (\ref{wvo}) and the properties of the WZW zero mode, the following 
results are obtained for the semiclassical WZW vertex operator: 

\vs .4cm
\newpage
\ni \un{Semiclassical unitarity of $g (\T,\bz,z)$ and Lie group elements}
\vs .3cm

Extreme semiclassical
unitarity of the WZW vertex operator
\bs
\be
g (\T,\bz,z) = \exp [ -  2i L^{ab}_g ( \bQ_a^-(\bz) + \bQ_a^+ (\bz)) \T_b ]
G(\T)   \exp [ 2 i L^{ab}_g ( Q_a^-(z) + Q_a^+(z) )   \T_b ]  
+ \oko
\ee
\be
g^{\dagger} (\T,\bz,z) g(\T,\bz,z) = \one + \oko 
\ee
\es 
follows from the extreme semiclassical unitarity (\ref{scu})
of the WZW zero mode, or from the
corresponding property of $\gp$ and $\gm$.

Moreover, we saw in (\ref{gfr}) and (\ref{wav})
 that the WZW zero modes satisfy the group 
multiplication and group integration laws in the semiclassical limit. 
Because the classical limit of the WZW vertex operator is the WZW zero mode
\be
g(\T,\bz,z) = G (\T) + \cO (k^{-1/2})
\ee
the same multiplication and integration laws are  
then obtained for the classical limits of  the products and averages of the
WZW vertex operators,   
\bs
\be
\eqalign{
g(\T^1,\bz_1,z_1)_{\a_1} & {}^{\b_1} 
g(\T^2,\bz_2,z_2)_{\a_2}{}^{\b_2} \cr  
& =\sum_{\T_k , \a_k, \b_k }  
\bC_{\a_1\a_2}{}^{\a_k}(\T^1 \T^2 \T^k) C^{\b_1\b_2}{}_{\b_k}(\T^1 \T^2 \T^k)  
g(\T^k,\bz_2,z_2)_{\a_k}{}^{\b_k} +\cO (k^{-1/2}) \cr }   
\ee
\be
\langle 0| g(\T^1,\bz_1,z_1) \cdots  g(\T^n,\bz_n,z_n) | 0 \rangle
   = \int {\rm d} \G \, \G (\T^1)  \cdots \G(\T^n) + \cO (k^{-1/2}) \pe  
\ee
\label{gma} \es
It is therefore consistent to identify 
the classical limit of both $g(\T,\bz,z)$ and  $G(\T)$ as  
the classical unitary group element $\G(\T)$ in irrep $\T$ of $g$, 
\bs
\be
g(\T,\bz,z) = \G (\T) + \cO (k^{-1/2})
\ee
\be
G(\T) = \G (\T) + \cO (k^{-1/2}) 
\ee
\be
\langle 0 | (\cdots)  | 0 \rangle = \int \rd \G (\cdots ) + \cO (k^{-1/2}) \pe 
\ee
\es 
To complete the classical results (\ref{gma}), the full semiclassical WZW 
OPE's and averages are given below. 

In the same way, the classical limits of the primary states
(\ref{aps}) of affine $(g\ti g)$
\be
\psi_\a{}^\b (\T) = g(\T,0,0)_\a{}^\b | 0 \rangle  = \G (\T)_a{}^\b |0 \rangle 
+\cO (k^{-1/2})  
\ee
are proportional to the classical group elements. This fact was first observed
in Ref.\cit{43}.  

\vs .4cm
\newpage
\ni \un{Semiclassical WZW OPE's}
\vs .3cm

The full OPE of two semiclassical WZW vertex operators is 
\be 
\eqalign{
g & (\T^1,\bz_1,z_1)_{\a_1}{}^{\b_1}   
g (\T^2,\bz_2,z_2)_{\a_2}{}^{\b_2}  \cr 
 & =  \sum_{\T_k , \a_k ,  \b_k   \atop   \r_1 ,\s_1}
 | z_{12}|^ {2[\D^g (\T^k) - \D^g(\T^1) -\D^g (\T^2)]} 
D_{\r_1 \a_2 \b_k}^{\s_1 \b_2 \a_k} (\T^1 \T^2 \T^k)  \cr
& \ti  \left\{ 
 g (\T^k,\bz_2,z_2 )_{\a_k}{}^{\b_k}  
\d_{\a_1}^{\r_1}   \d_{\s_1}^{\b_1}    
   \phantom{ \sum_{r=0}^{\infty} {z_{12}^{r+1} \over (r+1) !} }  
\right. \cr
& +  \sum_{r=0}^{\infty} {z_{12}^{r+1} \over (r+1) !}   
2L^{ab}_g :   g (\T^k,\bz_2,z_2)_{\a_k}{}^{\b_k}    \pa_2^r\J_b (z_2) : 
 \d_{\a_1}^{\r_1}   (\T^1_a)_{\s_1}{}^{\b_1}    
\cr  
& - \sum_{r=0}^{\infty} {\bz_{12}^{r+1} \over (r+1) !}   
 2L^{ab}_g :   g (\T^k,\bz_2,z_2)_{\a_k}{}^{\b_k}    \bpa_2^r \bJ_b (\bz_2) : 
(\T^1_a)_{\a_1}{}^{\r_1}   \d_{\s_1}^{\b_1}
\cr  
&  + \sum_{r,s=0}^{\infty} { z_{12}^{r+s+2} \over (r+s+2) !} 
      4 L^{ab}_g L^{cd}_g : g (\T^k,\bz_2,z_2)_{\a_k}{}^{\b_k}  
\pa_2^r [ \pa_2^s\J_b(z_2)\J_c(z_2)] :  
\d_{\a_1}^{\r_1}(\T^1_d\T^1_a)_{\s_1}{}^{\b_1}
 \cr  
&   +  \sum_{r,s=0}^{\infty} { \bz_{12}^{r+s+2} \over (r+s+2) !} 
      4 L^{ab}_g L^{cd}_g : g (\T^k,\bz_2,z_2)_{\a_k}{}^{\b_k}  
\bpa_2^r [ \bpa_2^s \bJ_b(\bz_2) \bJ_c(\bz_2)] :  
(\T^1_a\T^1_d)_{\a_1}{}^{\r_1}\d_{\s_1}^{\b_1}
 \cr  
&  \left. -  \sum_{r,s=0}^{\infty} {z_{12}^{r+1} \over (r+1) !} 
{\bz_{12}^{s+1} \over (s+1) !} 
 4 L^{ab}_g L^{cd}_g :   
g (\T^k,\bz_2,z_2)_{\a_k}{}^{\b_k}  \pa_2^r\J_b (z_2) \bpa_2^s \bJ_c (\bz_2): 
(\T^1_d)_{\a_1}{}^{\r_1} 
(\T^1_a)_{\s_1}{}^{\b_1}
 \right\} 
\cr  
 & \;\;\;\;\;\;\;\;\;\;\;\;\;\;\;\;\; 
+ \okt \cr}  
\ee 
where $D$ in (\ref{gfr}) is quadratic in the Clebsch-Gordan coefficients
$C$. It is observed that these OPE's have trivial monodromy, as they should,
with leading semiclassical terms which are the WZW primary fields themselves.

\vs .4cm
\ni \un{Semiclassical WZW correlators}
\vs .3cm

The semiclassical WZW correlators are 
\bs
\be
A_g (\T,\bz,z) = A_g^- (\T,\bz) A_g^+ (\T,z) 
=\langle 0 | g(\T^1,\bz_1,z_1) \cdots  g(\T^n,\bz_n,z_n) | 0\rangle 
\phantom{
\langle 0 |   g(\T^n,\bz_n,z_n) | 0\rangle}  
\ee
\be
= \left[ 1 + 2 L^{ab}_g \sum_{i <j}^n\T_a^i \T_b^j  \ln \bz_{ij} \right] 
\langle 0 |  G (\T^1) \cdots  G (\T^n)  | 0 \rangle   
\left[ 1 + 2 L^{ab}_g \sum_{i <j}^n\T_a^i \T_b^j  \ln z_{ij} 
\right] + \okt 
\ee 
\be
= \left[ 1 + 2 L^{ab}_g \sum_{i <j}^n\T_a^i \T_b^j  \ln \bz_{ij} \right] 
I_g^n 
\left[ 1 + 2 L^{ab}_g \sum_{i <j}^n\T_a^i \T_b^j  \ln z_{ij} 
\right] + \okt 
\ee
\be
= \left[ 1 + 4 L^{ab}_g \sum_{i <j}^n \T_a^i \T_b^j \ln |z_{ij}| \right] 
I_g^n 
+ \okt 
\ee
\label{wzc} \es
wher $I_g^n$ is the Haar integral in (\ref{hin}).
The result (\ref{wzc}), which is the central check on the results of this
paper, is the known \cit{42}  form of the 
 semiclassical WZW correlators. The simplicity of the result is due to the
fact that, once again, the normal-ordered terms in (\ref{wvo}) fail
to contribute to the correlators at this order of the semiclassical expansion. 
It is clear that these correlators
have trivial monodromy when one $z$ goes around another, and, moreover, the
chiral and antichiral intrinsic monodromies (\ref{imn}) have cancelled, so
that the intrinsic monodromies of $A_g$ are trivial. 

\section{Conclusions}

Supplementing the discussion of Moore and Reshetikhin \cit{25} and 
others [26-41] we have given
a new  semiclassical nonabelian vertex operator construction of the 
chiral and antichiral primary fields (the chiral and antichiral
vertex operators) associated to  
WZW theory, and the nonchiral primary fields of WZW theory itself. 

The new nonabelian vertex operators were obtained as the explicit semiclassical 
solution of known [26,25]  operator differential equations for the chiral 
and antichiral primary
fields, and they are the natural nonabelian generalization of the
familiar abelian vertex operators \cit{23}: The new vertex operators
 involve only
the representation matrices $\T$ of Lie $g$, the currents $\J,\bJ$ of affine 
$(g \ti g)$ and the chiral
and antichiral zero modes $G_{\pm} (\T) $, and they reduce to the familiar 
abelian vertex operators in the limit of abelian algebras.  So far as we
have carried out the semiclassical expansion, it was seen that the zero modes
carry the full action of the quantum group, and moreover, we were able
to identify the classical
limit of the nonchiral WZW zero mode $G(\T) =\Gm (\T) \Gp (\T)$ as the
classical group element in irrep $\T$ of $g$.

Combining our results with those of Ref.\cit{25}, we computed the semiclassical
OPE's among the chiral and antichiral vertex operators, and among the 
nonchiral WZW vertex operators themselves. Moreover, it was verified that
the new vertex operators reproduce the known \cit{42} form of the
semiclassical affine-Sugawara conformal blocks and WZW correlators, and
connections with semiclassical crossing matrices \cit{42} and 
braid relations [30,40] were also discussed.

We finally note that semiclassical blocks and correlators are
also known \cit{42} for the coset constructions and a class of processes
in irrational conformal field theory. Consequently, the present work should 
be considered
as a first step toward finding the semiclassical nonabelian vertex operators 
and OPE's of these more general theories. 
\newpage
\section*{Acknowledgements}

For many helpful discussions, we would like to thank our colleagues:
A. Alekseev, L. Alvarez-Gaum\'e, J. de Boer, E. Kiritsis, N. Reshetikhin,
P. Roche, A. Sagnotti and S. Shatashvili.  
We also thank the theory group at CERN for hospitality and support during
the course of this work. Finally, MBH acknowledges the hospitality and 
support of Ecole Polytechnique and NO acknowledges
the hospitality and support of the Niels Bohr Institute. 

The work of MBH was supported in part by the Director, Office of
Energy Research, Office of High Energy and Nuclear Physics, Division of
High Energy Physics of the U.S. Department of Energy under Contract
DE-AC03-76SF00098 and in part by the National Science Foundation under
grant PHY90-21139. The work of NO was supported in part by the European
Community, Project ERBCHBG CT93 0273.

\end{document}